\documentclass[a4paper,onecolumn]{article}

\usepackage[margin=0.6in]{geometry}

\usepackage[utf8]{inputenc}

\usepackage{graphicx}
\usepackage{subfig}
\usepackage{sidecap}
\usepackage{amssymb,amsfonts,amsmath}
\usepackage{nicefrac}
\usepackage{multicol}
\usepackage{color}
\usepackage{ulem} 

\title{Quadrupole shift cancellation using dynamic decoupling}

\begin{document}

\author{Ravid Shaniv, Nitzan Akerman, Tom Manovitz, Yotam Shapira and Roee Ozeri \\
Department of Physics of Complex Systems, The Weizmann Institute of Science, \\
Rehovot 7610001, Israel}

\maketitle

\begin{abstract}
We present a method that uses radio-frequency pulses to cancel the quadrupole shift in optical clock transitions. Quadrupole shifts are an inherent inhomogeneous broadening mechanism in trapped ion crystals, limiting current optical ion clocks to work with a single probe ion. Cancelling this shift at each interrogation cycle of the ion frequency allows the use of $N>1$ ions in clocks, thus reducing the uncertainty in the clock frequency by $\sqrt{N}$ according to the standard quantum limit. Our sequence relies on the tensorial nature of the quadrupole shift, and thus also cancels other tensorial shifts, such as the tensor ac stark shift. We experimentally demonstrate our sequence on three and seven $^{88}\mathrm{Sr}^{+}$ ions trapped in a linear Paul trap, using correlation spectroscopy. We show a reduction of the quadrupole shift difference between ions to $\approx20$ mHz's level where other shifts, such as the relativistic $2^{nd}$ order Doppler shift, are expected to limit our spectral resolution. In addition, we show that using radio-frequency dynamic decoupling we can also cancel the effect of 1$^{\mathrm{st}}$ order Zeeman shifts.
\end{abstract}

\twocolumn
%
%
Optical atomic clocks are one of the main achievements of quantum technology and provide some of the most accurate measurements to date, with applications ranging from basic science \cite{blatt2008new,chou2010optical,godun2014frequency} to technology \cite{ludlow2015optical,lisdat2016clock}. These clocks operate at optical frequencies, using laser light as a local oscillator (LO) that is locked to an atomic transition as a frequency reference. Large efforts are made to reduce LO noise as well as frequency shifts originating from coupling of the atomic reference to its surroundings \cite{ludlow2015optical}.\par
%
%
The stability of optical clocks depends primarily on the coupling between the atomic frequency reference and its environment. According to the standard quantum limit (SQL), the Signal to Noise Ratio (SNR) of a single experimental interrogation improves as $\sqrt{N}$ when interrogating $N>1$ atomic references simultaneously. Many identical atomic frequency references are used for example in optical lattice clocks, where $10^{5}$ or more atoms are interrogated at each experimental cycle \cite{campbell2017fermi}. \par
%
%
Another class of optical clocks is based on narrow optical transition in trapped ions. These clocks offer many advantages such as excellent experimental control, long coherence times, a long trapping lifetime and fast experimental duty-cycle for clock operation. In these clocks, the frequency reference is a narrow optical transition in an atomic ion trapped in a electromagnetic trap \cite{wineland1998experimental,ludlow2015optical}. In most such clocks, however, at least one of the clock transition levels has an electronic charge distribution that deviates from perfect spherical symmetry by a leading quadrupole term. This electronic charge distribution couples to any electric field gradient at the ion's position, and leads to a frequency shift of the clock's transition \cite{itano2000external}. When a single ion is used as the frequency reference, the shift arises primarily from trap DC electric field gradients, and therefore compromises the clock stability due to fluctuations in the electric potential. Typical values for the quadrupole shift range between $10-100 \text{ Hz}$ which is much larger than the clock linewidth ($1 \text{ Hz}$ or below). When multiple ions are trapped and form a crystal, the electric field gradient on each ion results from the crystal charge distribution in addition to the trap electrodes. Thus, the quadrupole shift for multiple ions in a chain is inhomogenous. As a result, most ion optical clocks work with a single ion frequency reference, limiting their averaging rate significantly.\par
%
%
Several techniques were developed in order to mitigate the quadrupole shift \cite{dube2005electric,itano2000external,margolis2004hertz}. In these methods the results of several ion frequency interrogations with different experimental parameters (e.g. different orientation of the quantization magnetic field) are averaged, and the quadrupole shift contribution vanishes. However, in the single ion case this comes at the price of lower duty cycle that might lead to degradation of the clock stability through the Dick effect \cite{quessada2003dick}. In the case of multiple ion clocks, these methods prove even less useful due to the inhomogeneous nature of the quadrupole shift.\par
%
%
Another way to overcome the quadrupole shift is choosing earth-alkaline-like ions with small quadrupole moment. Examples include $\text{In}^{+}$ and $\text{Al}^{+}$. However, these ions present different challenges, e.g. cooling transitions with UV wavelengths that are inaccessible to current laser technology, and therefore require cooling and detection either on the weak intercombination-line \cite{peik1994laser,eichenseer2003towards} or using logic spectroscopy \cite{schmidt2005spectroscopy}. \par
%
%
Here, we introduce a Ramsey-like spectroscopy scheme aimed at cancelling the quadrupole shift at each experimental interrogation, allowing for multiple-ion clock operation with no reduction in duty cycle. Our scheme is based on applying magnetic radio-frequency (RF) pulses producing a control time-dependent Hamiltonian during clock interrogation that takes advantage of the quadrupole shift symmetry and cancels its effect. It can be considered as a continuous Dynamical Decoupling (DD) scheme \cite{gordon2008optimal,fanchini2007continuously}. Furthermore, we show that using these pulses we can also cancel the first-order Zeeman shift. \par
%
%
We now turn to describe our scheme. Consider an excited level in a clock transition with a total spin of $J>\frac{1}{2}$, and an associated magnetic moment $\boldsymbol{\mu}_{z}=\gamma \textbf{J}_{z}$ interacting with a local magnetic field $\textbf{B} = B_{z}\hat{\textbf{z}}$. This excited state spin evolves according to the free Hamiltonian $\mathcal{H}_{\text{m}} = \hbar\gamma B_{z}\textbf{J}_{z}$. In the presence of an external gradient of the electric field across the ion position, an additional local quadrupole energy term is added: $\mathcal{H}_{\text{q}}=\hbar Q_{J}\left(\textbf{J}^{2}-3\textbf{J}^{2}_{z}\right)$, where $Q_{J}$ contains the atomic level's quadrupole moment, the gradient of the electric field and some geometric factors \cite{itano2000external}. Our strategy for cancelling the quadrupole moment relies on the identity $\textbf{J}^{2}-\left(\textbf{J}^{2}_{x}+\textbf{J}^{2}_{y}+\textbf{J}^{2}_{z}\right) = 0$. Elimination of the quadrupole shift is therefore possible by appropriate rotations of the spin $J$ operators. \par
To that aim, we now add a control RF drive resulting in the Hamiltonian term $\mathcal{H}_{\text{c}}=2\hbar \Omega\left(t\right)\cos\left(\omega_{\text{RF}}t-\phi\right)\textbf{J}_{x}$, where $\omega_{\text{RF}}=\frac{1}{\hbar}\gamma B_{z} +\delta\left(t\right)$ is close to the Larmor frequency, $\Omega\left(t\right)$ is the multi-level Rabi frequency and $\phi$ is the RF field phase. Moving to the interaction picture with respect to the drive and within the rotating wave approximation, we obtain the interaction Hamiltonian,
\begin{multline}
\mathcal{H}_{int}/\hbar=\delta\left(t\right)\textbf{J}_{z}+Q_{J}\left(\textbf{J}^{2}-3\textbf{J}^{2}_{z}\right)+\\
\Omega\left(t\right)\left[\cos\left(\phi\right)\textbf{J}_{x}+\sin\left(\phi\right)\textbf{J}_{y}\right].
\label{HamiltonianInt1}
\end{multline}
We assume that we can switch the RF drive on and off such that $\Omega\left(t\right)$ alternates between two values, $\Omega_{0}\gg Q_{J},\delta\left(t\right)$ and $0$, and $\phi\in\left[0,2\pi\right]$ is a controlled parameter. In the following we assume that $\delta\left(t\right)$ varies slowly on the ion's interrogation time scale, such that we can approximate it as constant, $\delta\left(t\right)\approx\delta$. When the RF field is switched off for time $\tau$, the spin state evolves freely according to the evolution operator
\begin{equation}
F\left(\tau\right)=\exp\left[i\left( \delta\tau\textbf{J}_{z}+Q_{J}\tau\left(\textbf{J}^{2}-3\textbf{J}^{2}_{z}\right)\right)\right].
\label{free_evolution}
\end{equation}
When short drive pulses are used, on time scales of $\tau\sim\frac{\pi}{\Omega_{0}}$, evolution due to the Hamiltonian terms $\delta \textbf{J}_{z}+Q_{J}\left(\textbf{J}^{2}-3\textbf{J}^{2}_{z}\right)$ can be neglected, and the evolution is then approximated as
\begin{multline}
\Pi^{\cos\left(\phi\right)x+\sin\left(\phi\right)y}\left(\Omega_{0}\tau\right)= \\
\exp\left[i \Omega_{0}\tau\left(\cos\left(\phi\right)\textbf{J}_{x}+\sin\left(\phi\right)\textbf{J}_{y}\right)\right].
\label{short_pulse_evolution}
\end{multline}
Here, we introduced the pulse time evolution operator $\Pi^{d}\left(\chi\right)$, where $d$ is the direction of the vector $\left(\cos\left(\phi\right),\sin\left(\phi\right)\right)$ in the $xy$ plane, and $\chi$ is the RF pulse area.
When longer pulse times are used, $\tau\gg\frac{\pi}{\Omega_{0}}$, the evolution takes a different form. Assuming for example $\phi=\frac{\pi}{2}$, the free evolution part in Eq. \ref{HamiltonianInt1} can be written as the sum of two terms:
$Q_{J}\left(\textbf{J}^{2}-\frac{3}{2}\left(\textbf{J}^{2}_{z}+\textbf{J}^{2}_{x}\right)\right)$ and $\delta\textbf{J}_{z}+Q_{J}\left(-\frac{3}{2}\left(\textbf{J}^{2}_{z}-\textbf{J}^{2}_{x}\right)\right)$. While the first term commutes with the control operator $\Omega_{0}\textbf{J}_{y}$, the latter does not, and is averaged out by the continuous operation of the control. As a result, this continuous pulse evolution becomes,
\begin{multline}
\Pi^{y}\left(\Omega_{0}\tau\right)=\exp\left[i\left(Q_{J}\tau(\textbf{J}^{2}-\frac{3}{2}(\textbf{J}^{2}_{z}+\textbf{J}^{2}_{x}))+\Omega_{0}\tau\textbf{J}_{y}\right)\right],
\label{long_pulse_evolution}
\end{multline}
where we used the same notation as for short pulses. This notation is chosen to match the geometric interpretation of the angular momentum operators as the generators of rotations, by which $\Pi^{x}\left(\pi\right)$ is a rotation of $\pi$ around the $x$ direction. \par
%
%
\begin{figure}[!t]
\includegraphics[width=0.5\textwidth]{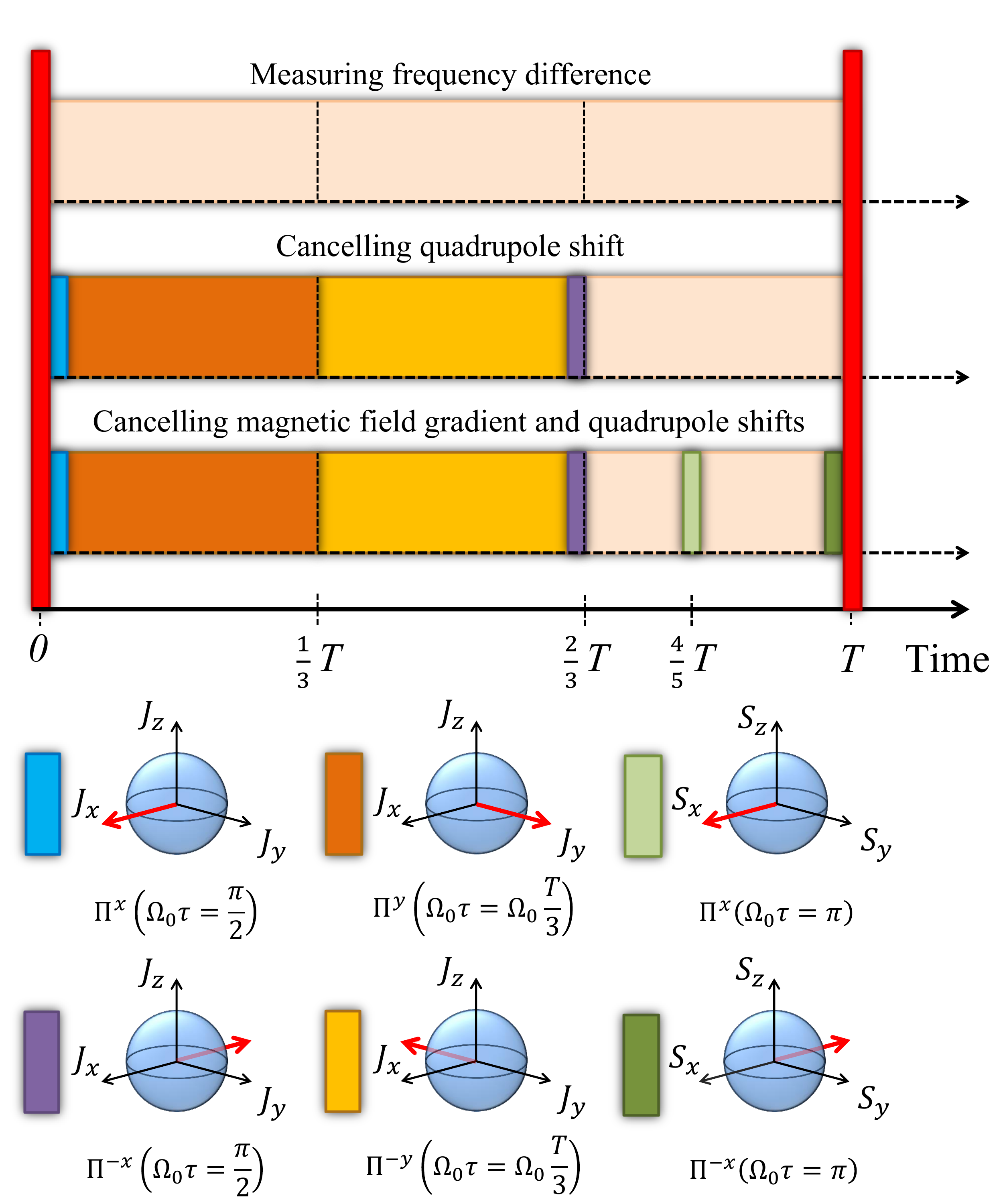}
\caption{\textbf{Experimental sequences scheme.} The sequence pulses are shown in different colors for different RF phases, different pulse times and different RF frequencies. In addition, a geometric representation legend for the angular momentum operators direction for each pulse. $J_{i}$ represents pulses resonant with the excited large spin manifold whereas $S_{i}$ represents pulses resonant with the ground state. Red operations mark the opening and closing optical Ramsey $\frac{\pi}{2}$ pulses.}
\label{experimentan_seq}
\end{figure}
Ramsey spectroscopy incorporating the DD sequence for cancelling the quadrupole shift is illustrated in Fig. \ref{experimentan_seq}. To match our experiment we describe our method explicitly using the example of a quadrupole allowed transition $\left|5S_{\frac{1}{2}},m_{S}\right\rangle\leftrightarrow\left|4D_{\frac{5}{2}},m_{D}\right\rangle$ of an earth alkaline ion, however we stress that our method is valid for any choice of J. Here, $m_{S}$ and $m_{D}$ are the corresponding magnetic quantum numbers, and the overall Ramsey interrogation time is $T$. The sequence begins with initializing all ions in $\left|5S_{\frac{1}{2}},m_{S}\right\rangle$. Next, an optical Ramsey $\frac{\pi}{2}$ pulse takes all ions to the equal superposition $\frac{1}{\sqrt{2}}\left(\left|5S_{\frac{1}{2}},m_{S}\right\rangle+\left|4D_{\frac{5}{2}},m_{D}\right\rangle\right)$. The DD operation begins with a RF rotary echo \cite{solomon1959rotary} sequence for a $\frac{2}{3}T$ interval, meaning the operation of $\Pi^{x}\left(\frac{\pi}{2}\right)$ rotating the state $\left|4D_{\frac{5}{2}},m_{D}\right\rangle$ from an eigenstate of $J_{z}$ with the eigenvalue $\hbar m_{D}$ to an eigenstate of $J_{y}$ with the same eigenvalue. Next, a $\Pi^{y}\left(\Omega_{0}\frac{T}{3}\right)$ is applied for time $\frac{1}{3}T$. During this operation, since the excited state manifold is in a $J_{y}$ eigenstate, and since $\left[Q_{J}(\textbf{J}^{2}-\frac{3}{2}(\textbf{J}^{2}_{z}+\textbf{J}^{2}_{x})),J_{y}\right]=0$, the evolution only results in a phase factor. This phase is global in the excited state manifold, but appears as a phase between the clock transition levels  that depends on $\Omega_{0}$, and must be corrected. Therefore, we apply a $\Pi^{-y}\left(\Omega_{0}\frac{T}{3}\right)$ operation, such that the contribution of the $\Omega_{0}\textbf{J}_{y}$ term in Eq. \ref{long_pulse_evolution} to the phase is reversed. The next step of the sequence is operating with $\Pi^{-x}\left(\frac{\pi}{2}\right)$. This pulse restores the ions' excited state to $\left|4D_{\frac{5}{2}},m_{D}\right\rangle$, an eigenstate of $Jz$. Finally, the state is left to freely evolve under $F\left(\frac{T}{3}\right)$ for the remained $\frac{T}{3}$ time. Therefore, the free evolution under the Hamiltonian in Eq. \ref{HamiltonianInt1} with $\Omega\left(t\right)=0$ results in an additional phase factor. We summarize the DD sequence evolution operator as:
\begin{equation}
F\left(\frac{T}{3}\right)\Pi^{-x}\left(\frac{\pi}{2}\right) \Pi^{-y}\left(\frac{\Omega_{0}T}{3}\right)\Pi^{y}\left(\frac{\Omega_{0}T}{3}\right)\Pi^{x}\left(\frac{\pi}{2}\right).
\label{sequence_operators}
\end{equation}
This evolution is exactly identical to

$\exp\left[i\frac{\delta T}{3\hbar}\textbf{J}_{z}+\frac{Q_{J}T}{\hbar}\left(\textbf{J}^{2}-\left(\textbf{J}^{2}_{x}+\textbf{J}^{2}_{y}+\textbf{J}^{2}_{z}\right)\right)\right]=\exp\left[i\frac{\delta T}{3\hbar}\textbf{J}_{z}\right]$

(see supplementary material), and the quadrupole shift term is eliminated. Finally, a second optical Ramsey $\frac{\pi}{2}$ pulse followed by state detection closes the Ramsey experiment. The residual frequency shift arising from the non-commuting terms neglected in Eq. \ref{long_pulse_evolution} scales as $p^{3}$, in the case where $p=\delta=Q_{J}$ (see supplementary material). As an example for $p=2\pi\times 10\text{ Hz}$ and $\Omega_{0}=2\pi\times 50\text{ kHz}$ the residual frequency shift would be smaller than $40 \mathrm{ \mu Hz}$. \par
%
%
\begin{figure}[!t]
\subfloat[]{\includegraphics[width=0.14\textwidth]{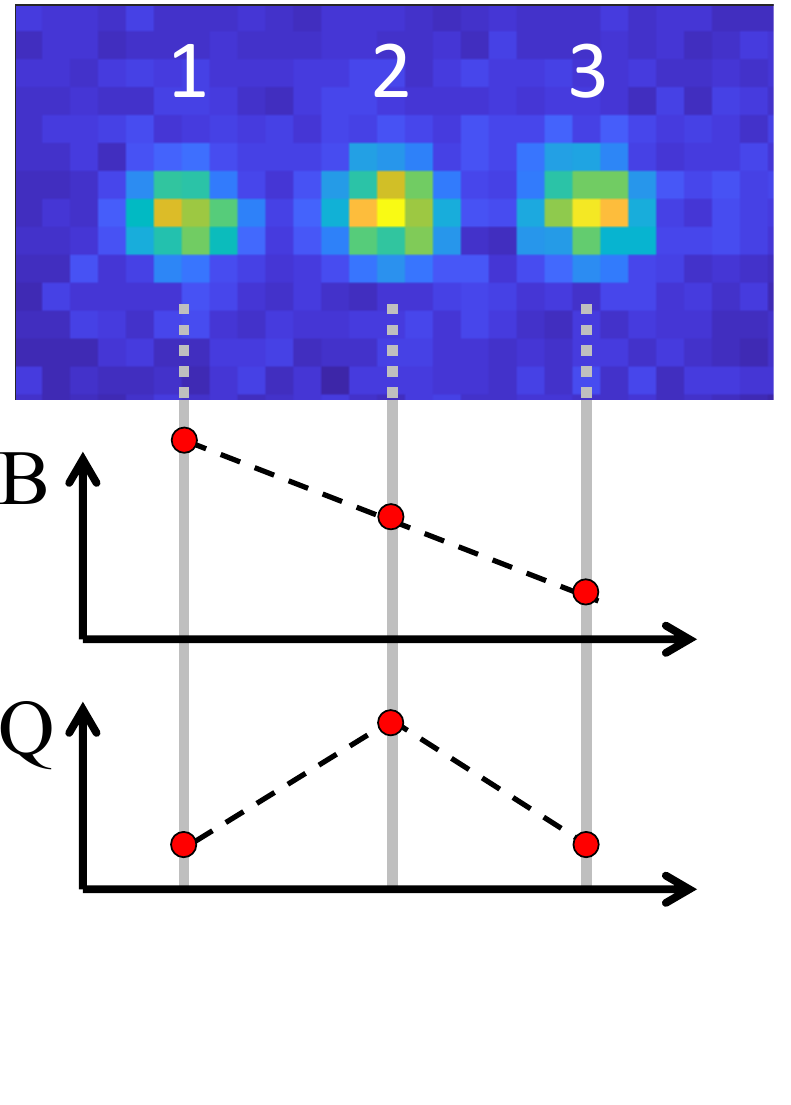}}
\subfloat[]{\includegraphics[width=0.305\textwidth]{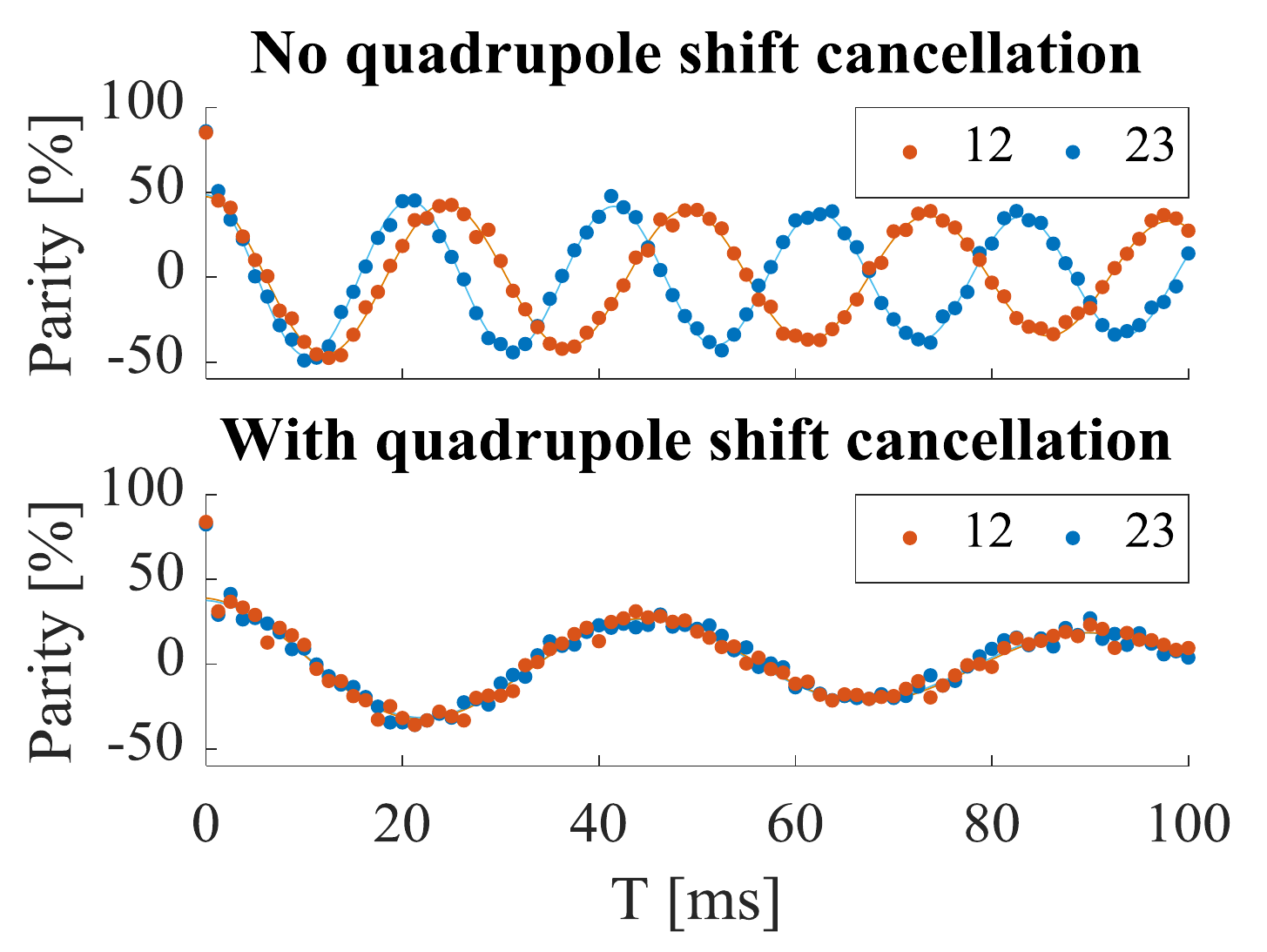}}
\hfill
\subfloat[]{\includegraphics[width=0.43\textwidth]{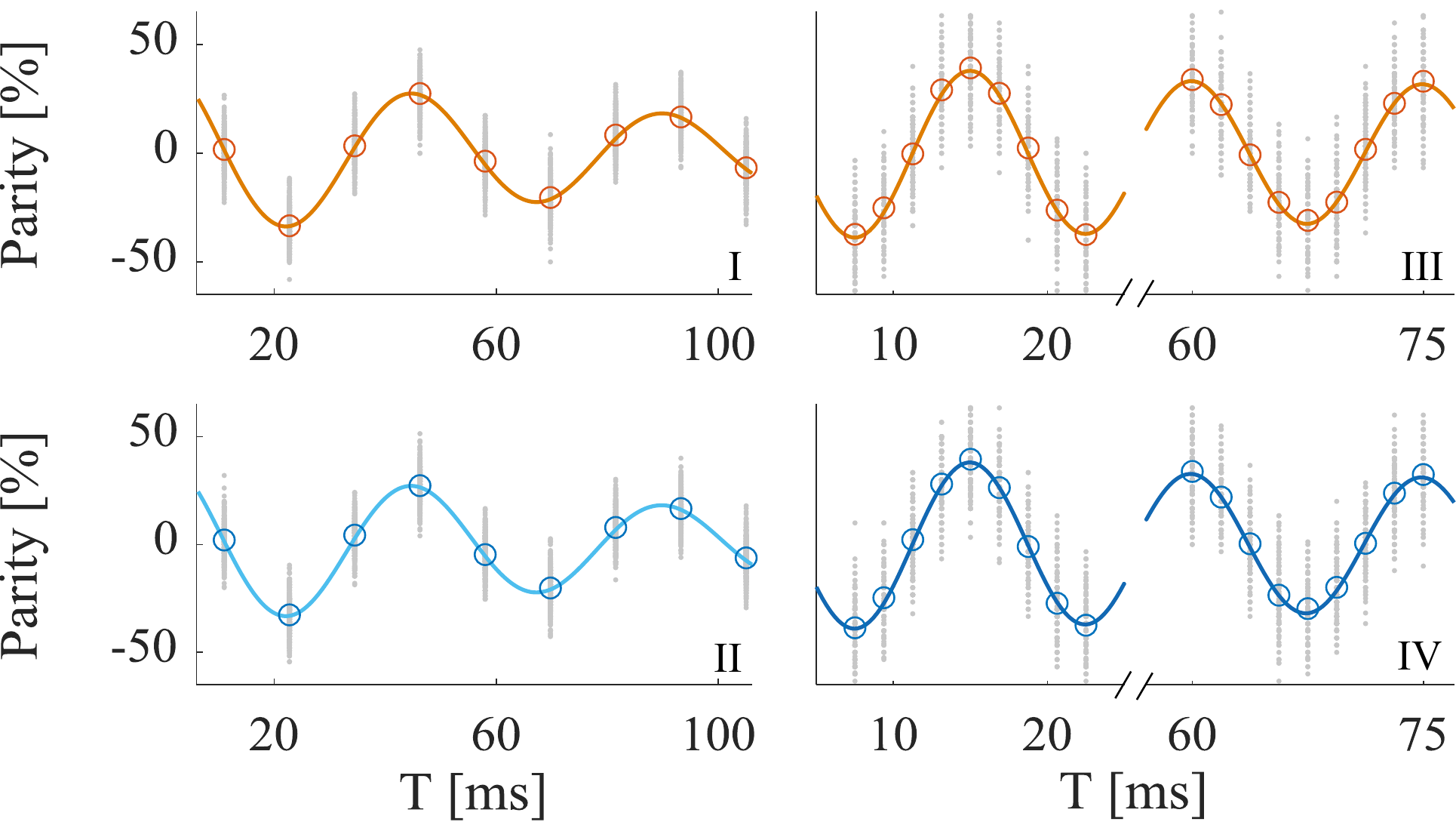}}
\caption{\textbf{Quadrupole shift cancellation results.} \textbf{(a)} Three ions image on an EMCCD camera, along with position-dependent magnetic field (M) and quadrupole shift (Q). \textbf{(b)} Parity oscillations of ions (1,2) and (2,3) in orange and blue circles. Upper and lower plots correspond to Ramsey and Quadrupole shift cancellation experimental sequences (see figure \ref{experimentan_seq}) respectively. The measured relative quadrupole shift is $\sim3.6\text{ Hz}$. Here the chosen superposition was $\frac{1}{\sqrt{2}}\left(\left|5S_{\frac{1}{2}},\text{-}\frac{1}{2}\right\rangle+\left|4D_{\frac{5}{2}},\text{-}\frac{3}{2}\right\rangle\right)$. Upper and lower plots exhibit different mean parity frequency Due to partial cancelation of Zeeman shifts in the DD sequence. \textbf{(c)} Quadrupole cancellation parity fringes with longer integration time. Frequencies were estimated by averaging frequencies from hundreds of minutes long Ramsey experiments (Grey dots). (I,II) Parity fringes of (1,2) and (2,3) pairs respectively for the above superposition after integration of 20.5 hours. (III,IV) Parity fringes of (1,2) and (2,3) pairs respectively for $\frac{1}{\sqrt{2}}\left(\left|5S_{\frac{1}{2}},\text{-}\frac{1}{2}\right\rangle+\left|4D_{\frac{5}{2}},\frac{1}{2}\right\rangle\right)$ superposition, after integration of 14 hours. Empty circles are the average of all the data (grey) points and and solid lines are maximum likelihood fits to the data, both exhibit similar frequencies agreement.}
\label{Quad_canc_experiment_results}
\end{figure}.\par
We verified the cancellation of the quadrupole shift in an experiment with a three $^{88}\mathrm{Sr}^{+}$ ion crystal trapped in a linear Paul trap. The optical quadrupole-allowed transition was addressed using a $\leq20\text{ Hz}$ linwidth 674 nm laser, and the ions' state was detected using state-selective fluorescence detection on an EMCCD camera. More details of the experimental apparatus can be found in \cite{akerman2015universal,TomThesis2016}. The axial trapping frequency was $1.5 \text{ MHz}$, resulting in a differential quadrupole shift between chain-edge and chain-middle ions of a few Hz for the $\left|4D_{\frac{5}{2}},m_{D}\right\rangle$ excited state.

To avoid laser phase noise and ambient magnetic field drifts, we took advantage of the fact that the quadrupole shift is different between different ions, and used correlation spectroscopy to infer the frequency difference between transitions in different ions \cite{chwalla2007precision, chou2011quantum}. Here, at the end of the Ramsey sequence described above the parity of the state, i.e. P(DD)+P(SS)-P(SD)-P(DS) where $P$ denotes probability and $S$ $\left(D\right)$ denotes states in the $\left|5S_{\frac{1}{2}},m_{s}\right\rangle$ $\left(\left|4D_{\frac{5}{2}},m_{D}\right\rangle\right)$ manifold, was measured. Assuming full decoherence of the single ion local oscillator, this parity signal will oscillate at the absolute value of the frequency difference between the two ions, with half the Ramsey fringe contrast.

In order to measure the reduction in the quadrupole shift, we applied a magnetic field gradient, creating a large Zeeman shift difference between adjacent ions. Considering a chain of $N$ ions, indexed from 1 to $N$ along the trap axis, we denote by $\omega_{i,j}=\left|\omega^{Q}_{i,j}+\omega^{M}_{i,j}\right|$ the parity fringe frequency of ions $i$ and $j$, where $\omega^{Q}_{i,j}$ and $\omega^{M}_{i,j}$ are its contributions due to quadrupole shift and magnetic field difference respectively. Note that $\omega_{i,j}=\omega_{j,i}$. However, due to the fact that the quadrupole shift and Zeeman shift are symmetric and anti-symmetric with respect to the chain center (see Fig. \ref{Quad_canc_experiment_results}a) we can write $\omega^{Q}_{i,j}=\omega^{Q}_{N+1-i,N+1-j}$ and $\omega^{M}_{i,j}=-\omega^{M}_{N+1-i,N+1-j}$, and therefore we can conclude that $\left|\omega_{i,j} - \omega_{N+1-j,N+1-j}\right|=\left|2\omega_{i,j}^{Q}\right|$. Therefore, cancelling the quadrupole shift means that the parity oscillation of ions ($i$,$j$) and ($N+1-j$,$N+1-j$) coincide.

Our experimental results for three ions are shown in Fig. \ref{Quad_canc_experiment_results}. As seen, without quadrupole shift cancellation the frequency shift of $\sim3.6\text{ Hz}$ between ion pairs (1,2) and (2,3)is clearly observed. With our DD sequence applied the parity oscillations of the two pairs seem to coincide.  A more quantitative measurement is shown In Fig. \ref{Quad_canc_experiment_results}c. Here parity fringes frequency for ion pairs (1,2) and (2,3) with superposition $\frac{1}{\sqrt{2}}\left(\left|5S_{\frac{1}{2}},\text{-}\frac{1}{2}\right\rangle+\left|4D_{\frac{5}{2}},\text{-}\frac{3}{2}\right\rangle\right)$ are $22.146\pm0.024\text{ Hz}$ and $22.168\pm0.024\text{ Hz}$ respectively, and therefore agree to within one standard deviation. For the superposition $\frac{1}{\sqrt{2}}\left(\left|5S_{\frac{1}{2}},\text{-}\frac{1}{2}\right\rangle+\left|4D_{\frac{5}{2}},\frac{1}{2}\right\rangle\right)$, Parity fringes frequency for ion pairs (1,2) and (2,3) are $66.722\pm0.034\text{ Hz}$ and $66.768\pm0.034\text{ Hz}$ respectively, and they agree to 1.3 standard deviation, corresponding to relative quadrupole shift cancellation at the level of $1\times10^{-16}$ with respect to the optical transition frequency,  limited by our statistical uncertainty.\par
\begin{figure}[!t]
\includegraphics[width=0.5\textwidth]{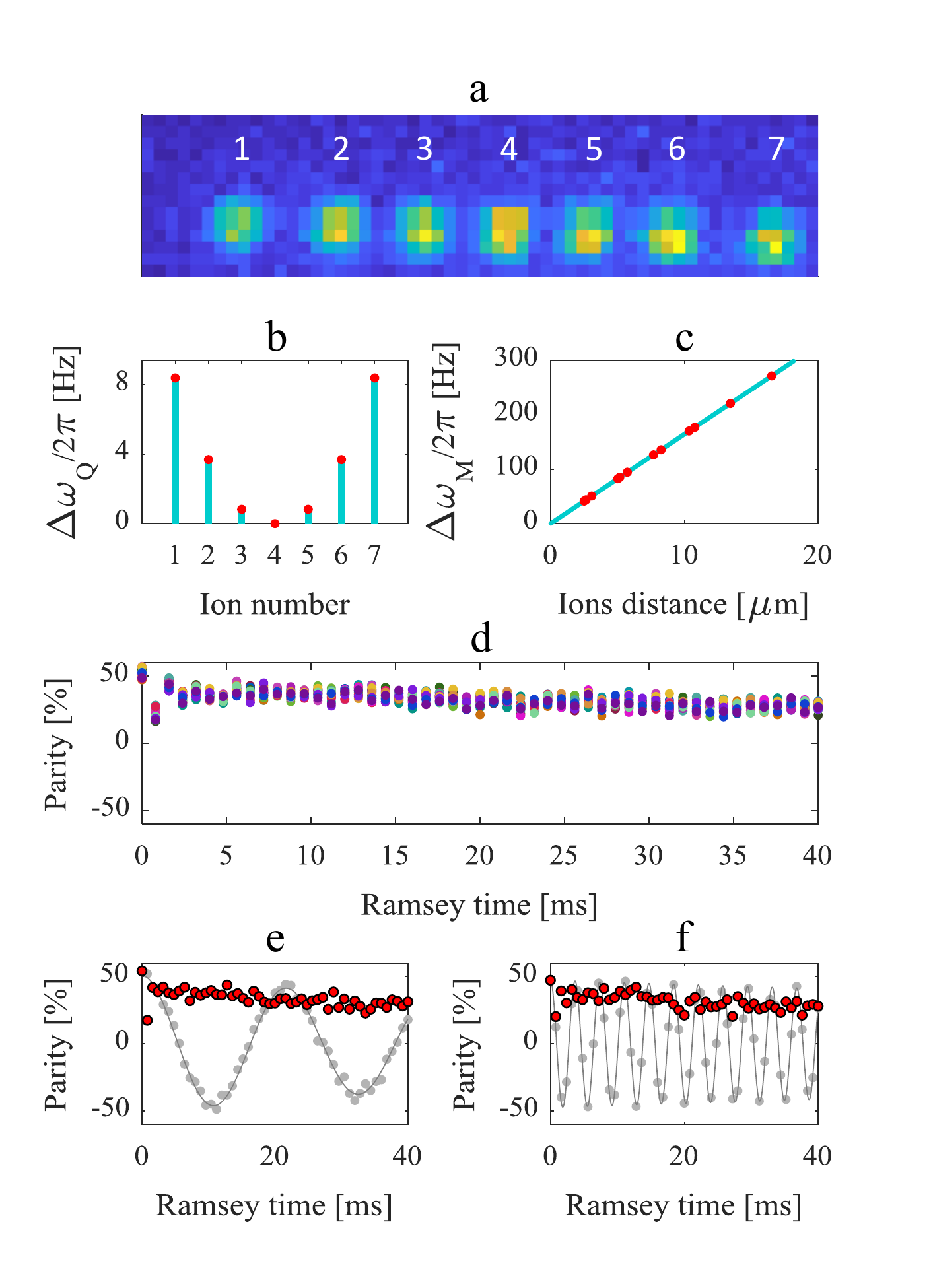}
\caption{\textbf{7 ions quadrupole shift and magnetic gradient cancellation experimental results.} \textbf{(a)} 7 ion chain imaged on a EMCCD camera. \textbf{(b)} Relative quadrupole shift ($\Delta\omega_{Q}$) measurement. The shift (red circles) is relative to ion number 4, and it is given by $\frac{1}{2}\left|\left|\omega_{i,4}\right|-\left|\omega_{8-i,4}\right|\right|$ for the $i$'th ion frequency. \textbf{(c)} Magnetic field gradient shift ($\Delta\omega_{M}$) measurement. Ion distance is calculated from trap parameters, and the shift (red circles) is calculated as $\frac{1}{2}\left|\left|\omega_{i,j}\right|+\left|\omega_{8-i,8-j}\right|\right|$. A linear fit (turquoise solid line) yields $\approx16.4\text{ Hz}/\mu m$. Standard deviations for both (b) and (c) range from $60$ to $130\text{ mHz}$ and therefore are too small to be presented. \textbf{(d)} Quadrupole and magnetic field gradient shifts cancellation experimental results, of all 21 correlations. \textbf{(e,f)} Correlations (1,2) and (1,7) and their fit as examples for inhomogeneous shifts measurement experiment (grey circles and solid lines) respectively, along with their counterparts in the cancellation experiment.}
\label{7_ion_experiment}
\end{figure}\par
%
%
The experiment reported above was performed over the course of several weeks and not consecutively. Between one experiment and the next drifts in the frequency-matching between symmetric correlations (with respect to the chain center) were observed, up to $230\text{ mHz}$. Although the source of these drifts is not fully identified, we believe they are not residual quadrupole shifts. First, the quadrupole shift in our lab is stable and changes by less than 10\% over days. In addition, our DD cancellation method was identically repeated in the different experimental realizations, and the residual quadrupole shift for our experimental parameters is expected to be less than $2\text{ mHz}$ (see supplementary material).  Therefore, residual quadrupole shifts cannot explain run-to-run variations in the symmetric correlations of $100's\text{ mHz}$. We note, however, that in these experiments we did not use mechanical shutters. The observed shifts could originate due to differential light shifts from leaks of laser light during the Ramsey interrogation time. \par
%
In addition to the quadrupole shift cancellation, Additional DD RF pulses can cancel the remaining $1^{\text{st}}$ order Zeeman shifts from the last part of free evolution \cite{akerman2015universal}. Here we apply an additional RF $\pi$ pulse that flips the two Zeeman states in the $\left|5S_{\frac{1}{2}},m_{S}\right\rangle$ manifold during the final wait time as shown in Fig. \ref{experimentan_seq}. The pulse time is determined taking into account the magnetic response of both excited and ground states. Fig. \ref{7_ion_experiment} summarizes the experimental results for cancelling both $1^{\text{st}}$ order Zeeman and quadrupole shifts on a 7-ion chain. As seen, following the cancellation sequence no apparent parity oscillations are observed, regardless of the correlation pair examined. We can then conclude that $1^{\text{st}}$ order Zeeman and quadrupole shift are largely reduced. The reduction in contrast is due to noisy magnetic field spectral components that overlap with our sequence. It can be reduced by appropriately optimizing the RF intensity which in our case would be lowering it (see supplementary material). \par
%
%
Aside from the electric quadrupole interaction and position dependent magnetic field, relativistic Doppler effect can also cause a position dependent frequency shift due to time-dilation in the moving ion reference frame \cite{chou2010optical,ludlow2015optical}. Since our trap has a significant axial micro-motion gradient \cite{navon2013addressing} we expect different ions to exhibit different measured frequency  due to the associated 2nd order Doppler shift. We estimate these shifts along the trap axis to be in the $50\text{ mHz}$ scale. We therefore do not expect to verify the quadrupole shift cancellation to better than this scale. We would also point out that by cancelling the quadrupole and $1^{\text{st}}$ order Zeeman shifts on the ions, our sequence can be used for measuring this relativistic effect. For clock application this inhomogeneous shift can be reduced greatly by a more symmetric trap design \cite{keller2018controlling}. \par
%
%
Resonant RF drive applied on the excited state Zeeman separation is an off-resonance drive to the ground state Zeeman manifold. As a result, the latter's level spacing would shift due to ac Stark effect. Taking as example a RF Rabi frequency on the ground state Zeeman transition of $\Omega_{S}\approx2\pi\times150\text{ kHz}$ and detuning of $\delta_{S}\approx3.5\text{ MHz}$ the ground state levels will be shifted by $\frac{\Omega_{S}^{2}}{\delta_{S}}=2\pi\times6.5\text{ kHz}$. This is a large unwanted shift originating from the cancellation sequence. In order to mitigate this shift, one could first choose lower Rabi frequency, and therefore reduce this shift quadratically. An additional strategy uses the fact that the ground state levels are shifted symmetrically around the zero magnetic field energy. Two approaches can be taken. First, one could average measurements done on two transitions with different $m$ levels. For example, $\left|5S_{\frac{1}{2}},\frac{1}{2}\right\rangle\leftrightarrow\left|4D_{\frac{5}{2}},\frac{3}{2}\right\rangle$ and $\left|5S_{\frac{1}{2}},\text{-}\frac{1}{2}\right\rangle\leftrightarrow\left|4D_{\frac{5}{2}},\text{-}\frac{3}{2}\right\rangle$. This technique is widely used in current optical atomic clocks to eliminate first order Zeeman shifts, but has the disadvantage of lowering the sampling rate and therefore increasing frequency uncertainties due to the Dick effect. Second, additional RF pulses applied on the ground state Zeeman manifold can eliminate any phase accumulation originating in a shift symmetric to the zero magnetic field energy, including magnetic field shifts and the ac shift mentioned above (see supplementary material). \par
%
%
The DD sequence presented above cancels any shift having angular momentum dependence of $\mathcal{H}_{q}$. In addition to the quadrupole shift, the tensor polarizability has this form, and therefore our sequence can eliminate this shift as well. Although this shift is usually small for ion-based optical clocks, in neutral atoms trapped in an optical lattice these shifts can be significant, and impose stringent demands on the lattice beam polarization \cite{ludlow2015optical}. \par
%
%
To conclude, in this work we proposed and experimentally demonstrated a method to cancel quadrupole shifts in trapped ions systems. This shift is the main obstacle for the use of many ions in an ion-based optical atomic clock, and therefore cancelling this shift allows for better signal to noise and in turn more accurate and precise clock operation. Our sequence is based on RF pulses, which are cheap and experimentally easy to implement. In addition, we show that using these RF pulses sequences position dependent magnetic field effects can be cancelled as well. By using correlation spectroscopy we experimentally showed reduction of inhomogeneity from electric quadrupole interaction to $1\times10^{-16}$, where other effects might limit our resolution. We demonstrate cancellation of both quadrupole and magentic field shifts between ions on a seven-ion chain, proving that our scheme is applicable for multiple ion chain clocks.

In parallel to our studies, a similar method that incorporates dynamic-decoupling techniques for the cancellation of quadrupole shifts was investigated \cite{RetzkerReference}.

This work was supported by the Crown Photonics Center, ICore-Israeli excellence center
circle of light, the Israeli Science Foundation, the Israeli Ministry of Science Technology
and Space, the Minerva Stiftung and the European Research Council (consolidator grant 616919-Ionology).

\bibliographystyle{unsrt}
\bibliography{quad_canc_ref}

\part*{Supplementary material}

\section*{DD sequence evolution computation}
The DD sequence for quadrupole cancellation reads

$F\left(\frac{T}{3}\right)\Pi^{-x}\left(\frac{\pi}{2}\right) \Pi^{-y}\left(\frac{\Omega_{0}T}{3}\right)\Pi^{y}\left(\frac{\Omega_{0}T}{3}\right)\Pi^{x}\left(\frac{\pi}{2}\right)$.
In order to compute the superposition phase accumulated through the sequence, we first write explicitly
\begin{multline}
    \Pi^{-y}\left(\frac{\Omega_{0}T}{3}\right)\Pi^{y}\left(\frac{\Omega_{0}T}{3}\right)= \\ \exp\left[i\frac{1}{\hbar}\left(Q_{J}\frac{T}{3}(\textbf{J}^{2}-\frac{3}{2}(\textbf{J}^{2}_{z}+\textbf{J}^{2}_{x}))-\Omega_{0}\frac{T}{3}\textbf{J}_{y}\right)\right] \\ \exp\left[i\frac{1}{\hbar}\left(Q_{J}\frac{T}{3}(\textbf{J}^{2}-\frac{3}{2}(\textbf{J}^{2}_{z}+\textbf{J}^{2}_{x}))+\Omega_{0}\frac{T}{3}\textbf{J}_{y}\right)\right]= \\
    \exp\left[i \frac{Q_{J}T}{\hbar}(\frac{2}{3}\textbf{J}^{2}-(\textbf{J}^{2}_{z}+\textbf{J}^{2}_{x}))\right]
\end{multline}
We now note that the operators $\Pi^{x}\left(\frac{\pi}{2}\right)$ and $\Pi^{-x}\left(\frac{\pi}{2}\right)$ act as a rotation of $\frac{\pi}{2}$ angle around the $x$ axis. Therefore, the operation of the first four operators is equivalent to the operation of the evolution operator
\begin{equation}
U_{1}=\exp\left[i\frac{Q_{J}T}{\hbar}(\frac{2}{3}\textbf{J}^{2}-(\textbf{J}^{2}_{y}+\textbf{J}^{2}_{x}))\right].
\end{equation} The final free evolution part is
\begin{equation}
U_{2}=\exp\left[i\frac{\delta T}{3\hbar} \textbf{J}_{z}+\frac{Q_{J}T}{\hbar}\left(\frac{1}{3}\textbf{J}^{2}-\textbf{J}^{2}_{z}\right)\right].
\end{equation}
$U_{1}$ and $U_{2}$ commute and therefore the total sequence evolution results in
\begin{multline}
U_{2}U_{1}= \\
\exp\left[i\frac{\delta T}{3\hbar} \textbf{J}_{z}+i\frac{Q_{J}T}{\hbar}\left(\textbf{J}^{2}-\left(\textbf{J}^{2}_{x}+\textbf{J}^{2}_{y}+\textbf{J}^{2}_{z}\right)\right)\right]= \\
\exp\left[i\frac{\delta T}{3\hbar} \textbf{J}_{z}\right]
\end{multline}
matching the result stated in the main text.

\section*{Fast magnetic noise}
\begin{figure}[!t]
    \centering
    \includegraphics[width=0.4\textwidth]{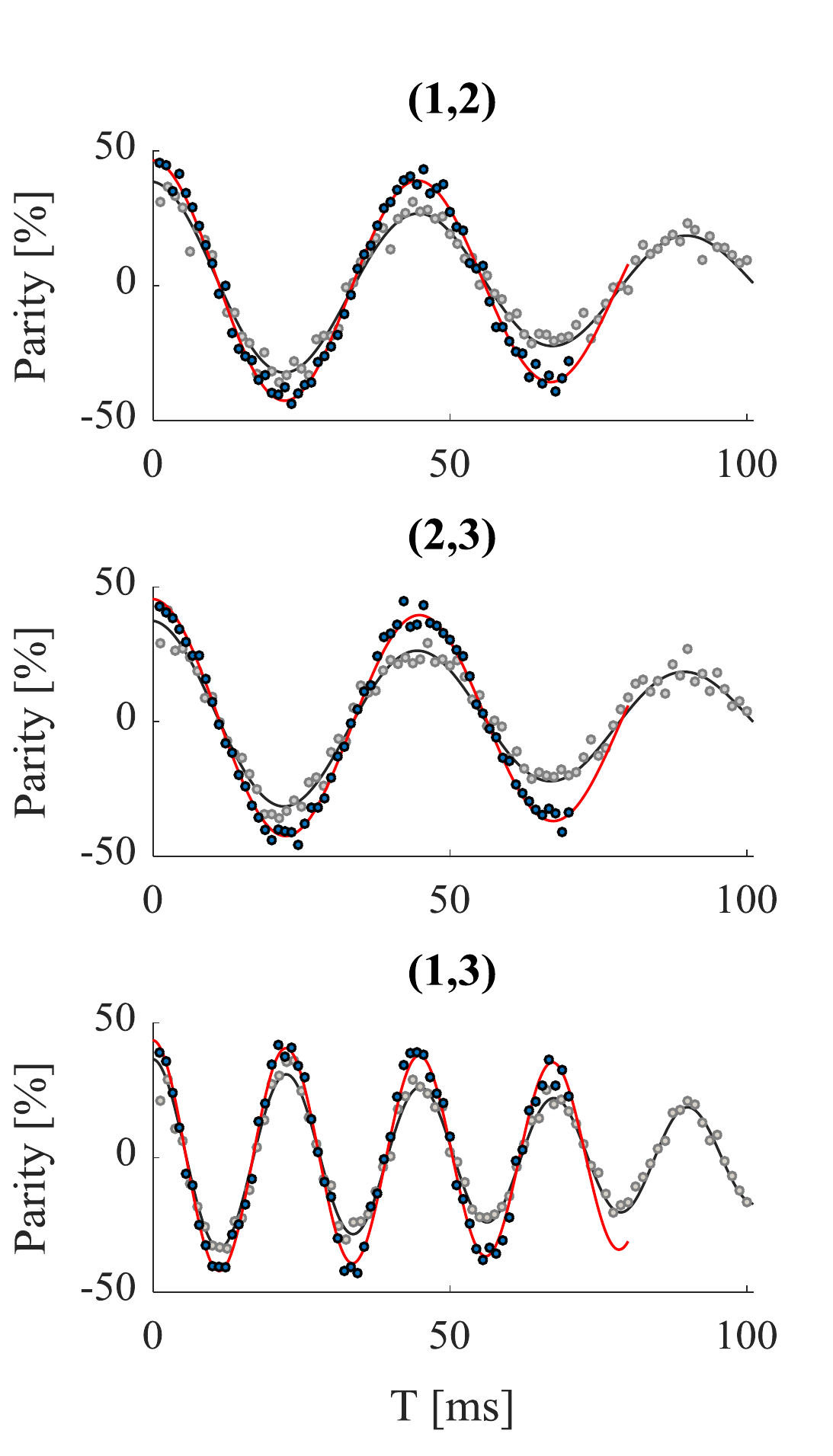}
    \caption{\textbf{Parity fringes with low and high RF power.} Each plot shows two parity fringes taken with the quadrupole cancellation sequence: one for Rabi frequency $\Omega_{0}\approx2\pi\times90\text{ kHz}$ (grey circles) with the parity fit to the theoretical model (lack solid line), and one for Rabi frequency $\Omega_{0}\approx2\pi\times47\text{ kHz}$ (dark blue circles) along with its corresponding fit to the theoreical model (red solid line). The plots from top to bottom show the parity fringes for the (1,2),(2,3),(1,3) ion pairs, corresponding to the ion numbers in figure 2 in the main text.}
    \label{Low_High_Amp_comparison}
\end{figure}
\begin{figure}[!t]
    \centering
    \includegraphics[width=0.4\textwidth]{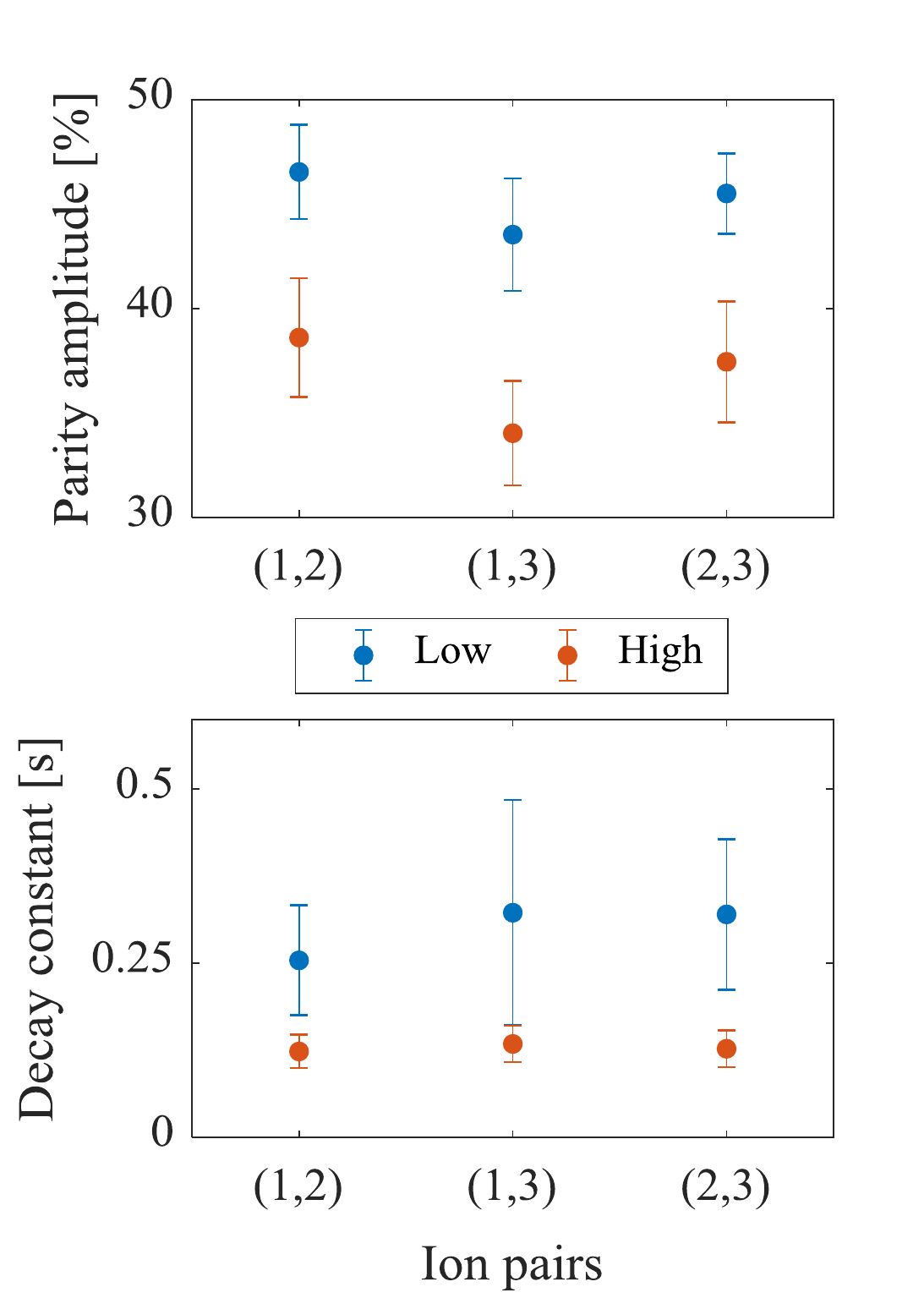}
    \caption{\textbf{Comparison between fringe amplitude and decay time for high and low RF power.} \textbf{(Top, Bottom)} Maximum likelihood estimation for the parameter $a,b$ in equation \ref{fringe_fitting_model} respectively, with error-bars marking 95\% confidence interval. Blue (Orange) circles mark the estimated parameter for low (high) Rabi frequency.}
    \label{Low_High_Amp_fit_parameters}
\end{figure}

In our experiment, lower Ramsey fringe contrast was observed in experiments cancelling the quadrupole shift with respect to experiments measuring it. We believe it is due to fast magnetic noise components that were not compensated. These noise components overlap with our dynamical decoupling sequence spectrum. When choosing a lower RF Rabi frequency, the contrast improves. The results of experimental comparison between two DD pulse Rabi frequencies are shown in Fig. \ref{Low_High_Amp_comparison}. Here we compared the parity oscillations for two different RF field amplitudes. The Rabi frequencies were approximately $90$ and $47$ kHz for high and low amplitudes, respectively. The parity fringes were fitted to the expression
\begin{equation}
    a\times\exp\left[-\frac{t}{b}\right]\cos\left[2\pi\times c t\right]
    \label{fringe_fitting_model}
\end{equation}
where $a,b,c$ are fit parameters corresponding to the fringe amplitude, fringe decay constant and fringe frequency. Comparison between the amplitude and decay fit parameters for all three of the ion pair correlations is shown in Fig. \ref{Low_High_Amp_fit_parameters}. From the plot it is evident that with lower RF power the contrast increases by roughly 10\% compared to the higher power. In addition, the decay time is larger in the lower power experiment. We then conclude that the lower contrast depends on the RF power, and can be minimized. Therefore, it is not a limitation to our sequence. In the experiment shown in Fig. 2c (III,IV) in the main text, the lower RF amplitude was used.

\section*{Radio-frequency ac stark-shifts and magnetic field shifts cancellation}
%
%
Due to the fact that the first part of our quadrupole cancellation sequence uses continuous RF drive resonant with the excited state Zeeman manifold separation, the ground state Zeeman manifold is ac stark-shifted. We define the magnetic response in angular frequency per magnetic field of the ground state as $\chi_{g}$ and of the excited state as $\chi_{e}$. The magnetic quantum numbers of the ground and excited states are denoted as $m_{g}$ and $m_{e}$ respectively, and for a given RF power we define $\Omega_{g}$ and $\Omega_{e}$ as the ground and excited state magnetic Rabi frequencies. \\
%
%
Applying the sequence mentioned in Eq. 5 in the main text results in a final phase shift between the excited and ground state that does not include laser phase drifts. In the first part of the sequence, $\frac{2}{3}$ of the total interrogation time $T$, the excited state accumulates phase due to quadrupole shift contributions only (since the magnetic field drifts are averaged out by the drive and the drive phase cancels due to the rotary echo). However, the ground state accumulated phase has two contributing terms: magnetic field drifts part $\frac{2}{3}\times\chi_{g}m_{g}BT$ where $B$ is the ambiant magnetic field, and ac stark shift term $\pm\frac{2}{3}\times\frac{\Omega_{g}^{2}}{2\delta}T$, where $\delta=\left(\chi_{e}-\chi_{g}\right)B$ and the shift's sign is the sign of $m_{g}$. Since these shifts are dependent differently on $\Omega_{g}$ and $B$, they cannot cancel each other in a robust way. In the second part of the experiment, the last $\frac{1}{3}T$, the ions are free-evolving according to Eq. 5 in the main text. While this free evolution cancels the quadrupole shift phase entirely, magnetic field phase continues to accumulate in both excited and ground states. This phase between ground and excited states takes the form of $\frac{1}{3}\left(-\chi_{g}m_{g}+\chi_{e}m_{e}\right)BT$. In total, the final superposition not taking into account laser drifts can be written as $\frac{1}{\sqrt{2}}\left(\exp{i\phi}\left|g\right\rangle+\left|e\right\rangle\right)$ where $\phi$ is given by
\begin{equation}
    \phi=\chi_{g}m_{g}BT-\frac{1}{3}\chi_{e}m_{e}BT\pm\frac{2}{3}\times\frac{\Omega_{g}^{2}}{2\delta}T.
    \label{unwanted_phase}
\end{equation}
%
%
These shifts can be eliminated with the insertion of echo pulses on the ground state manifold. These pulses switches the sign phase accumulation in the first and third terms in Eq. \ref{unwanted_phase}.
\paragraph*{Magnetic field cancellation.} Disregarding the last term in Eq. \ref{unwanted_phase}, we can see that plugging in $^{88}\text{Sr}^{+}$ parameters $\chi_{g}=2.802 \text{ MHz}/\text{G}$ and
$\chi_{e}=1.68 \text{ MHz}/\text{G}$ and using the transition of $m_{g}=-\frac{1}{2}$ and $m_{e}=-\frac{3}{2}$, applying an echo pulse on the ground state manifold at some $\tau$ results in the phase:
\begin{eqnarray}
&&\phi=\chi_{g}m_{g}B\tau-\frac{1}{3}\chi_{e}m_{e}B\tau \nonumber \\ &&+\left(-\chi_{g}m_{g}B\left(T-\tau\right)-\frac{1}{3}\chi_{e}m_{e}B\left(T-\tau\right)\right) \nonumber \\
&&=2.802\left(-\frac{1}{2}\right)B\tau-\frac{1}{3}1.68\left(-\frac{3}{2}\right)B\tau \nonumber \\ &&+\left(-2.802\left(-\frac{1}{2}\right)B\left(T-\tau\right)-\frac{1}{3}1.68\left(-\frac{3}{2}\right)B\left(T-\tau \nonumber \right)\right)
\end{eqnarray}
Equating this phase to zero we get $\tau=\frac{4}{5}$, which corresponds to the echo time in the last sequence of figure 1 in the main text. A second ground-state echo is then applied just before the final optical Ramsey pulse in order to retrieve the initial ground level within the ground Zeeman manifold. \\
\paragraph*{Magnetic field and ac stark shift cancellation.} In this case, we again use two ground state echo pulses. The first pulse is used to cancel the light shift phase. To that aim, it should be applied at $\tau_{1}=\frac{1}{3}T$. This timing also cancels any magnetic phase accumulation during the first $\frac{2}{3}$ of the interrogation. Therefore, we are left with the magnetic field shifts from the $\frac{2}{3}T$ to $T$. Here, the pulse time should be chosen in order to null the remained phase
\begin{eqnarray}
&&\phi_{\text{remain}}=\left(\chi_{g}m_{g}-\chi_{e}m_{e}\right)B\left(\tau_{2}-\frac{2}{3}T\right)\nonumber \\
&&+\left(-\chi_{g}m_{g}-\chi_{e}m_{e}\right)B\left(T-\tau_{2}\right).
\end{eqnarray}
  Equating $\phi_{\text{remain}}=0$ gives the second pulse time
  \begin{equation}
      \tau_{2}=\frac{1}{6}\left(\frac{\chi_{e}m_{e}}{\chi_{g}m_{g}}+5\right)T. \nonumber
  \end{equation}
Depending on the chosen states, a situation in which $\tau_{2}>T$ might occur, for example in the choice of states above. In this case, instead of inserting a second $\pi$ pulse in the ground state manifold it should be inserted in the excited state manifold, and its time should be calculated appropriately. The above sequence is a minimal control sequence, requiring only two echo pulses. However, we comment that more pulses and more sophisticated echo sequences can be used, that can be more efficient in filtering out magnetic noise. \\
%
%
We also note here that since the ground-state ac stark-shift depends on $\Omega_{g}^{2}$, it is beneficial to reduce the RF intensity as much as possible. The optimal value should depend on both quadrupole shift magnitude, magnetic field magnitude and magnetic field power spectral density. For example, taking $\Omega_{e}=2\pi\times5\text{ kHz}$ corresponds to $\Omega_{g}$ with the same order of magnitude. Assuming the same parameters of $^{88}\text{Sr}^{+}$, and assuming $1^{\text{st}}$ order Zeeman shift  $\approx10\text{ Hz}$ at $50 \text{ Hz}$ frequency shift, the light shift will then be in the several Hz's level, which is easily reduced to the mHz level by echos as mentioned above.
\onecolumn

\section*{Deviations from ideal continuous DD}
We would like to estimate how well does the continuous DD approximation made in equation 4 in the main text describes the real evolution. The approximation states that the continuous DD evolution operator

$\exp\left[i\frac{T}{3}\left(\delta J_{z}+Q_{J}\left(J^{2}-3J_{z}^{2}\right)-\Omega_{0}J_{y}\right)\right]$ has a similar operation to the evolution operator

$\exp\left[i\frac{T}{3}\left(Q_{J}\left(J^{2}-\frac{3}{2}\left(J_{z}^{2}+J_{x}^{2}\right)\right)-\Omega_{0}J_{y}\right)\right]$. Instead of focusing on part of the DD evolution, we evaluate the difference between the actual and approximated entire continuous DD part of our sequence evolution. Here, the actual evolution takes the form of
\begin{multline}
    U_{act}\left(T\right)=\exp\left[-i\frac{\pi}{2}\boldsymbol{J}_{x}\right]\exp\left[i\frac{T}{3}\left(\delta \boldsymbol{\boldsymbol{J}}_{z}+Q_{J}\left(\boldsymbol{J}^{2}-3J_{z}^{2}\right)-\Omega_{0}\boldsymbol{J}_{y}\right)\right]\\
    \exp\left[i\frac{T}{3}\left(\delta \boldsymbol{J}_{z}+Q_{J}\left(\boldsymbol{J}^{2}-3\boldsymbol{J}_{z}^{2}\right)+\Omega_{0}\boldsymbol{J}_{y}\right)\right]
    \exp\left[i\frac{\pi}{2}\boldsymbol{J}_{x}\right]=\\
    \exp\left[i\frac{\Omega_{0} T}{3}\left(-\frac{\delta}{\Omega_{0}} \boldsymbol{J}_{y}+\frac{Q_{J}}{\Omega_{0}}\left(\boldsymbol{J}^{2}-3J_{y}^{2}\right)-\boldsymbol{J}_{z}\right)\right]\exp\left[i\frac{\Omega_{0} T}{3}\left(-\frac{\delta}{\Omega_{0}} \boldsymbol{J}_{y}+\frac{Q_{J}}{\Omega_{0}}\left(\boldsymbol{J}^{2}-3J_{y}^{2}\right)+\boldsymbol{J}_{z}\right)\right],
    \label{actual_DD_operator}
\end{multline}
and the approximated evolution is
\begin{equation}
    U_{app}\left(T\right)=\exp\left[i\frac{2T}{3}\left(Q_{J}\left(\boldsymbol{J}^{2}-\frac{3}{2}\left(\boldsymbol{J}_{x}^{2}+\boldsymbol{J}_{y}^{2}\right)\right)\right)\right]. \label{approximate DD operator}
\end{equation}
We define:
$\boldsymbol{V}=-\frac{\delta}{\Omega_{0}}\boldsymbol{J}_{y}-\frac{3Q_{J}}{2\Omega_{0}}\left(\boldsymbol{J}_{y}^{2}-\boldsymbol{J}_{x}^{2}\right)$ and $\boldsymbol{H}=\frac{Q_{J}}{\Omega_{0}}\left(\boldsymbol{J}^{2}-\frac{3}{2}\left(\boldsymbol{J}_{y}^{2}+\boldsymbol{J}_{x}^{2}\right)\right)$, and we note that $\boldsymbol{V}$ has zeros for its diagonal elements, while $\boldsymbol{H}$ is diagonal with diagonal elements different from zero. Using these definitions we rewrite the operators $U_{act}$ and $U_{app}$:
\begin{equation}
    U_{act}\left(T\right)=\exp\left[i\frac{\Omega_{0} T}{3}\left(\boldsymbol{V}+\boldsymbol{H}-\boldsymbol{J}_{z}\right)\right]\exp\left[i\frac{\Omega_{0} T}{3}\left(\boldsymbol{V}+\boldsymbol{H}+\boldsymbol{J}_{z}\right)\right],
    \label{actual V,H,Jz}
\end{equation}
\begin{equation}
    U_{app}\left(T\right)=\exp\left[i\frac{2\Omega_{0} T}{3}\boldsymbol{H}\right]. \label{approximate V,H,Jz}
\end{equation}

For a mutual eigenstate $\left|m\right\rangle$ of $\boldsymbol{J}^{2}$ and $\boldsymbol{J}_{z}$, we define $\left\langle m\right|U_{act}\left(T\right)\left|m\right\rangle=r_{act}\exp\left(i \phi_{act}\right)$ and $\left\langle m\right|U_{app}\left(T\right)\left|m\right\rangle=\exp\left(i \phi_{app}\right)$. The latter has magnitude of unity due to the fact that $\left|m\right\rangle$ is an eigenstate of $\boldsymbol{H}$. We denote the $\boldsymbol{H}$ eigenvalue corresponding to $\left|m\right\rangle$ by $h_{m}$. With this notation, we write $\phi_{app}=\frac{\Omega_{0}T}{3}\times2 h_{m}$

We now assume for simplicity that $\frac{\delta}{\Omega_{0}}=\frac{Q_{J}}{\Omega_{0}}=p\ll1$, and note that the operators $\boldsymbol{V},\boldsymbol{H}$ and the number $h_{m}$ are proportional to $p$. We now turn to prove that the elements that contribute to the phase $\phi_{act}-\phi_{app}$ with the lowest power in $p$ are proportional to $p^{3} \Omega_{0}T$.

\subsection*{Theoretical proof for $p^{3}$ scaling}
We examine the quantity $\left\langle m\right|U_{act}\left(T\right)U_{app}^{\dagger}\left(T\right)\left|m\right\rangle=r_{act}\exp\left(i\left(\phi_{act}-\phi_{app}\right)\right)=g\left(p,\Omega_{0},\right)+if\left(p,\Omega_{0},\right)$,  where $f$ and $g$ are real functions describing real and imaginary parts.

\subsubsection*{Reduction to imaginary part}

We claim that the lowest power of $p$ in a Taylor expansion of the phase $\phi_{act}-\phi_{app}$ is equal to the lowest power of $p$ in the Taylor expansion of the imaginary part $f$. Since $g\left(0,\Omega_{0}\right)=1$ and $f\left(0,\Omega_{0}\right)=0$ (see Eq. \ref{actual_DD_operator} and Eq. \ref{approximate DD operator}), we can write
$g\left(p,\Omega_{0}\right)=1+\left(\text{powers of }p\right)$, and $f\left(p,\Omega_{0}\right)=0+\left(\text{powers of }p\right)$. This leads to the conclusion that the lowest power of $p$ in the expansion of the phase of $\left\langle m\right|U_{act}U_{app}^{\dagger}\left(T\right)\left|m\right\rangle$, which is $\arctan\left(\frac{f\left(p,\Omega_{0}\right)}{g\left(p,\Omega_{0}\right)}\right)$, has the same lowest power of $p$ as $f\left(p,\Omega_{0}\right)$. Therefore, it is sufficient to prove that $f\left(p,\Omega_{0}\right)$ does not contain a linear or quadratic $p$ term.

\subsubsection*{$p^{3}$ scale}

We now expand $\left\langle m\right|U_{act}U_{app}^{\dagger}\left(T\right)\left|m\right\rangle$ in a Taylor series:
\begin{equation}
    \left\langle m\right|U_{act}U_{app}^{\dagger}\left|m\right\rangle=\sum_{n,k}\frac{1}{n!k!}\left(i\frac{\Omega_{0}T}{3}\right)^{n+k} \underset{T_{nk}}{\underbrace{\left\langle m\right|\left(\boldsymbol{V}+\boldsymbol{H}-h_{m}\boldsymbol{I}+\boldsymbol{J}_{z}\right)^{n}\left(\boldsymbol{V}+\boldsymbol{H}-h_{m}\boldsymbol{I}-\boldsymbol{J}_{z}\right)^{k}\left|m\right\rangle}}, \label{Taylor series}
\end{equation}
where $\boldsymbol{I}$ is the unity operator and where we defined $T_{nk}$ as a sum coefficient. $T_{nk}$ can be written as summation over terms proportional to
\begin{equation}
\left\langle m\right|...\left[\left(\boldsymbol{H}-h_{m}\boldsymbol{I}\right)^{l_{i}}\boldsymbol{J}_{z}^{j_{i}}\left(-\boldsymbol{J}_{z}\right)^{j'_{i}}\boldsymbol{V}^{v_{i}}\right]\left[\left(\boldsymbol{H}-h_{m}\boldsymbol{I}\right)^{l_{i+1}}\boldsymbol{J}_{z}^{j_{i+1}}\left(-\boldsymbol{J}_{z}\right)^{j'_{i+1}}\boldsymbol{V}^{v_{i+1}}\right]...\left|m\right\rangle, \label{T_{nk} terms}
\end{equation}
 where for each $i=1,2,3,...,n+k$ one and only one of $\{l_{i},j_{i},j'_{i},v_{i}\}\text{ is } 1$ and the rest are zeros, $\sum_{i}\left(j_{i}\right)\le n$,  $\sum_{i}\left(j'_{i}\right)\le k$  and $\sum_{i}\left(l_{i}+j_{i}+j'_{i}+v_{i}\right)=n+k$.

 We now prove two statements:

 \textbf{(*) All terms in which $\sum_{i}\left(v_{i}\right)=0$ cancel out.}

 \textbf{proof:}

 Since $\left(\boldsymbol{J}_{z}\right)$ and $\left(\boldsymbol{H}-h_{m}\boldsymbol{I}\right)$ commute, these terms can be written as:
 $\left\langle m\right|\boldsymbol{J}_{z}^{\sum_{i}j_{i}}\left(-\boldsymbol{J}_{z}\right)^{\sum_{i}j'_{i}}\left(\boldsymbol{H}-h_{m}\boldsymbol{I}\right)^{\sum_{i}l_{i}}\left|m\right\rangle$. In cases where $\sum_{i}l_{i}\ne 0$, the above expression equals $\left\langle m\right|\boldsymbol{J}_{z}^{\sum_{i}j_{i}}\left(-\boldsymbol{J}_{z}\right)^{\sum_{i}j'_{i}}\left|m\right\rangle\left(h_{m}-h_{m}\right)^{\sum_{i}l_{i}}=0$. In cases where $\sum_{i}l_{i}= 0$, the sum over all terms with $n+k=d$ takes the form $\sum_{n}\frac{n!}{\left(n\right)!\left(d-n\right)!}\left\langle m\right|\left(\boldsymbol{J}_{z}\right)^{n}\left(-\boldsymbol{J}_{z}\right)^{k}\left|m\right\rangle=\left\langle m\right|\left(\boldsymbol{J}_{z}-\boldsymbol{J}_{z}\right)^{d}\left|m\right\rangle=0$.

 \textbf{(**) Terms in which $\sum_{i}\left(v_{i}\right)=1$ always vanish.}

 \textbf{proof:}

 Due to the fact that $\left(\boldsymbol{V}\right)$ has only zeros as its diagonal elements, it transforms a state $\left|m\right\rangle$ to a state orthogonal to it. On the other hand, both $\left(\boldsymbol{H}-h_{m}\boldsymbol{I}\right)$ and $\boldsymbol{J}_{z}$ are diagonal in the $\left|m\right\rangle$ basis, and therefore transforms a state $\left|m\right\rangle$ to a state proportional to it. As a result, the state
 $\left(\boldsymbol{H}-h_{m}\boldsymbol{I}\right)^{a}\boldsymbol{J}_{z}^{b}\left(-\boldsymbol{J}_{z}\right)^{c}\boldsymbol{V}\left(\boldsymbol{H}-h_{m}\boldsymbol{I}\right)^{a'}\boldsymbol{J}_{z}^{b'}\left(-\boldsymbol{J}_{z}\right)^{c'}\left|m\right\rangle$, for integer $a,b,c,a',b',c'$, must be orthogonal to $\left|m\right\rangle$.

  Statements \textbf{(*)} and  \textbf{(**)} prove that all terms proportional to $p$ vanish.

  Only $T_{nk}$ terms in which $n+k$ is odd contribute to $f\left(p,\Omega_{0}\right)$. Therefore, we prove that terms proportional to $p^{2}$ cancel out for odd $n+k$. Terms proportional to $p^{2}$ are divided to three types:

  1 - $\sum_{i}v_{i}=0$ and $\sum_{i}l_{i}=2$

  2 - $\sum_{i}v_{i}=1$ and $\sum_{i}l_{i}=1$

  3 - $\sum_{i}v_{i}=2$ and $\sum_{i}l_{i}=0$

  Types 1 and 2 vanish due to the two statements \textbf{(*)} and  \textbf{(**)} above. We now deal with type 3. As is written above, only terms with odd $n+k$ contribute to $f\left(p,\Omega_{0}\right)$. We assume a term in $T_{nk}$ in which $n+k$ is odd and $\sum_{i}v_{i}=2$ and $\sum_{i}l_{i}=0$. We must have $\sum_{i}j_{i}+\sum_{i}j'_{i}=n+k-2$, and that means that either $\sum_{i}j_{i}$ is odd and $\sum_{i}j'_{i}$ is even, or $\sum_{i}j_{i}$ is even and $\sum_{i}j'_{i}$ is odd. In either case, a similar term with switched $\sum_{i}j_{i}$ and $\sum_{i}j'_{i}$ must appear in $T_{kn}$, and due to opposite sign, these terms will cancel in summation.

  \subsubsection*{Bound for the residual phase}

  We would now show that the phase $\phi_{act}-\phi_{app}$ has a dominant part proportional to $p^{3}\Omega_{0}T$. We will show it in an example for $J=\frac{5}{2}$ spin and for the state $\left|m=\frac{5}{2}\right\rangle$. We now look at all the imaginary terms in Eq. \ref{Taylor series}, meaning terms for which $n+k$ is odd, and are proportional to $p^{3}$. These terms must be in one of two forms:
  \begin{enumerate}
      \item $\left\langle m\right|\left(\boldsymbol{J}_{z}\right)^{g_{1}}\boldsymbol{V}\left(\boldsymbol{J}_{z}\right)^{g_{2}}\boldsymbol{V}\left(\boldsymbol{J}_{z}\right)^{g_{3}}\boldsymbol{V}\left(\boldsymbol{J}_{z}\right)^{g_{4}}\left|m\right\rangle$,
      \item $\left\langle m\right|\left(\boldsymbol{J}_{z}\right)^{g_{1}}\boldsymbol{V}\left(\boldsymbol{J}_{z}\right)^{g_{2}}\left(\boldsymbol{H}-h_{m}\boldsymbol{I}\right)\left(\boldsymbol{J}_{z}\right)^{g_{3}}\boldsymbol{V}\left(\boldsymbol{J}_{z}\right)^{g_{4}}\left|m\right\rangle$.
  \end{enumerate}
  Note that $g_{1}+g_{2}+g_{3}+g_{4}=n+k-3$. In addition, these terms must have even number of $\left(-\boldsymbol{J}_{z}\right)$ and even number of $\left(\boldsymbol{J}_{z}\right)$, because all terms that do not satisfy this condition cancel in summation with a corresponding term in $T_{kn}$.
  We now calculate each of these two terms for some choice of $g_{1},g_{2},g_{3},g_{4}$ for $\left|m=\frac{5}{2}\right\rangle$:

  We begin with type 1 term. The operator $\boldsymbol{V}$ couples a state $\left|m\right\rangle$ to the states $\left|m+1\right\rangle,\left|m+2\right\rangle,\left|m-1\right\rangle,\left|m-2\right\rangle$, with the coefficients $v_{m\rightarrow m+1},v_{m\rightarrow m+2},v_{m\rightarrow m-1},v_{m\rightarrow m-2}$ respectively.
  We can therefore compute:
  \begin{multline}
  \left\langle \frac{5}{2}\right|\left(\boldsymbol{J}_{z}\right)^{g_{1}}\boldsymbol{V}\left(\boldsymbol{J}_{z}\right)^{g_{2}}\boldsymbol{V}\left(\boldsymbol{J}_{z}\right)^{g_{3}}\boldsymbol{V}\left(\boldsymbol{J}_{z}\right)^{g_{4}}\left|\frac{5}{2}\right\rangle=\\
  \left(\frac{5}{2}\right)^{g_{1}+g_{4}}\left[\left(\frac{3}{2}\right)^{g_{2}}\left(\frac{1}{2}\right)^{g_{3}}V_{\frac{3}{2}\rightarrow\frac{5}{2}}V_{\frac{1}{2}\rightarrow\frac{3}{2}}V_{\frac{5}{2}\rightarrow\frac{1}{2}}+\left(\frac{1}{2}\right)^{g_{2}}\left(\frac{3}{2}\right)^{g_{3}}V_{\frac{1}{2}\rightarrow\frac{5}{2}}V_{\frac{3}{2}\rightarrow\frac{1}{2}}V_{\frac{5}{2}\rightarrow\frac{3}{2}}\right]
  \end{multline}
  Next, we compute type 2 term:
  \begin{eqnarray}
  \left\langle \frac{5}{2}\right|\left(\boldsymbol{J}_{z}\right)^{g_{1}}\boldsymbol{V}\left(\boldsymbol{J}_{z}\right)^{g_{2}}\left(\boldsymbol{H}-h_{\frac{5}{2}}\boldsymbol{I}\right)\left(\boldsymbol{J}_{z}\right)^{g_{3}}\boldsymbol{V}\left(\boldsymbol{J}_{z}\right)^{g_{4}}\left|\frac{5}{2}\right\rangle=\\
  \left(\frac{5}{2}\right)^{g_{1}+g_{4}}\left[V_{\frac{3}{2}\rightarrow\frac{5}{2}}V_{\frac{5}{2}\rightarrow\frac{3}{2}}\left(h_{\frac{3}{2}}-h_{\frac{5}{2}}\right)\left(\frac{3}{2}\right)^{g_{2}+g_{3}}+V_{\frac{1}{2}\rightarrow\frac{5}{2}}V_{\frac{5}{2}\rightarrow\frac{1}{2}}\left(h_{\frac{1}{2}}-h_{\frac{5}{2}}\right)\left(\frac{1}{2}\right)^{g_{2}+g_{3}}\right]
  \end{eqnarray}

  When we plug the numbers $V_{\frac{5}{2}\rightarrow\frac{3}{2}}=p (-i\frac{\sqrt{5}}{2})$, $V_{\frac{5}{2}\rightarrow\frac{1}{2}}=p (3\sqrt{\frac{5}{2}})$, $V_{\frac{3}{2}\rightarrow\frac{1}{2}}=p (i\sqrt{2})$, $h_{\frac{3}{2}}-h_{\frac{5}{2}}=-6p$ and $h_{\frac{1}{2}}-h_{\frac{5}{2}}=-9p$ we obtain:
  \begin{multline}
  \left\langle \frac{5}{2}\right|\left(\boldsymbol{J}_{z}\right)^{g_{1}}\boldsymbol{V}\left(\boldsymbol{J}_{z}\right)^{g_{2}}\boldsymbol{V}\left(\boldsymbol{J}_{z}\right)^{g_{3}}\boldsymbol{V}\left(\boldsymbol{J}_{z}\right)^{g_{4}}\left|\frac{5}{2}\right\rangle=\\
      p^{3}\left(-\frac{15}{2}\right)\left(\frac{5}{2}\right)^{g_{1}+g_{4}}\left[\left(\frac{3}{2}\right)^{g_{2}}\left(\frac{1}{2}\right)^{g_{3}}+\left(\frac{1}{2}\right)^{g_{2}}\left(\frac{3}{2}\right)^{g_{3}}\right]
  \end{multline}
  and
  \begin{multline}
  \left\langle m\right|\left(\boldsymbol{J}_{z}\right)^{g_{1}}\boldsymbol{V}\left(\boldsymbol{J}_{z}\right)^{g_{2}}\left(\boldsymbol{H}-h_{m}\boldsymbol{I}\right)\left(\boldsymbol{J}_{z}\right)^{g_{3}}\boldsymbol{V}\left(\boldsymbol{J}_{z}\right)^{g_{4}}\left|m\right\rangle=\\
      p^{3}\left(\frac{5}{4}\right)\left(\frac{5}{2}\right)^{g_{1}+g_{4}}\left[\left(-6\right)\left(\frac{3}{2}\right)^{g_{2}+g_{3}}+\left(-9\right)\left(\frac{1}{2}\right)^{g_{2}+g_{3}}\right]
  \end{multline}

  These calculations show that both terms have the same sign. In addition, we write that replacing each of the operators $\boldsymbol{V}$ and $\boldsymbol{H}-h_{\frac{5}{2}}\boldsymbol{I}$ with the operator $\left(h_{\frac{1}{2}}-h_{\frac{5}{2}}\right)\boldsymbol{I}$ yields a more negative number for any choice of $g_{1},g_{2},g_{3},g_{4}$:
  \begin{multline}
  \left\langle \frac{5}{2}\right|\left(\boldsymbol{J}_{z}\right)^{g_{1}}\left(h_{\frac{1}{2}}-h_{\frac{5}{2}}\right)\boldsymbol{I}\left(\boldsymbol{J}_{z}\right)^{g_{2}}\left(h_{\frac{1}{2}}-h_{\frac{5}{2}}\right)\boldsymbol{I}\left(\boldsymbol{J}_{z}\right)^{g_{3}}\left(h_{\frac{1}{2}}-h_{\frac{5}{2}}\right)\boldsymbol{I}\left(\boldsymbol{J}_{z}\right)^{g_{4}}\left|\frac{5}{2}\right\rangle=  p^{3}\left(-27\right)\left(\frac{5}{2}\right)^{g_{1}+g_{2}+g_{3}+g_{4}}
  \end{multline}

  This means that the sum
  \begin{equation}
  p^{3}\left(-27\right)\sum_{d\ge3\mbox{ odd}}\sum_{n=0}^{d}\frac{1}{n!\left(d-n\right)!}\left(i\frac{\Omega_{0}T}{3}\right)^{d}\left(\frac{5}{2}\right)^{d-3}
  \end{equation}

  places a conservative bound on the residual phase. This sum is easily calculated and is equal to
  \begin{equation}
    i p^{3} \left(\frac{2}{5}\right)^{3} \left(-27\right)\left(\sin\left(2\frac{\Omega_{0} T}{3}\times\frac{5}{2}\right)+\left(2\frac{\Omega_{0} T}{3}\times\frac{5}{2}\right)\right)
  \end{equation}
  which is the sum of a linear and a bounded oscillating terms, both scale as $p^{3}$.

  To conclude, we proved that the residual frequency difference, $f_{r}$, between the operation of actual DD operator and the approximated one enters to leading order as a linear combination of terms scaling as $\frac{\delta^{3}}{\Omega_{0}^{3}}\Omega_{0}$, $\frac{\delta^{2}Q_{J}}{\Omega_{0}^{3}}\Omega_{0}$, $\frac{\delta Q_{J}^{2}}{\Omega_{0}^{3}}\Omega_{0}$  and $\frac{Q_{J}^{3}}{\Omega_{0}^{3}}\Omega_{0}$, with an added bounded oscillating term.
  In addition, the statements \textbf{(*)} and  \textbf{(**)} also prove that $r_{act}$ deviates from 1 only as of $p^{2}$ to first order.

  \textbf{Note: For clarity, the proof assumed $\frac{\delta}{\Omega_{0}}=\frac{Q_{J}}{\Omega_{0}}=p$. A similar proof holds when using $p_{1}=\frac{\delta}{\Omega_{0}}$ and $p_{2}=\frac{Q_{J}}{\Omega_{0}}$, and the conclusion above is still derived.}
\newpage
\subsection*{numerical verification for $p^{3}$ scaling}

\begin{figure}[!h]
    \centering
    (a) \includegraphics[width=0.43\textwidth]{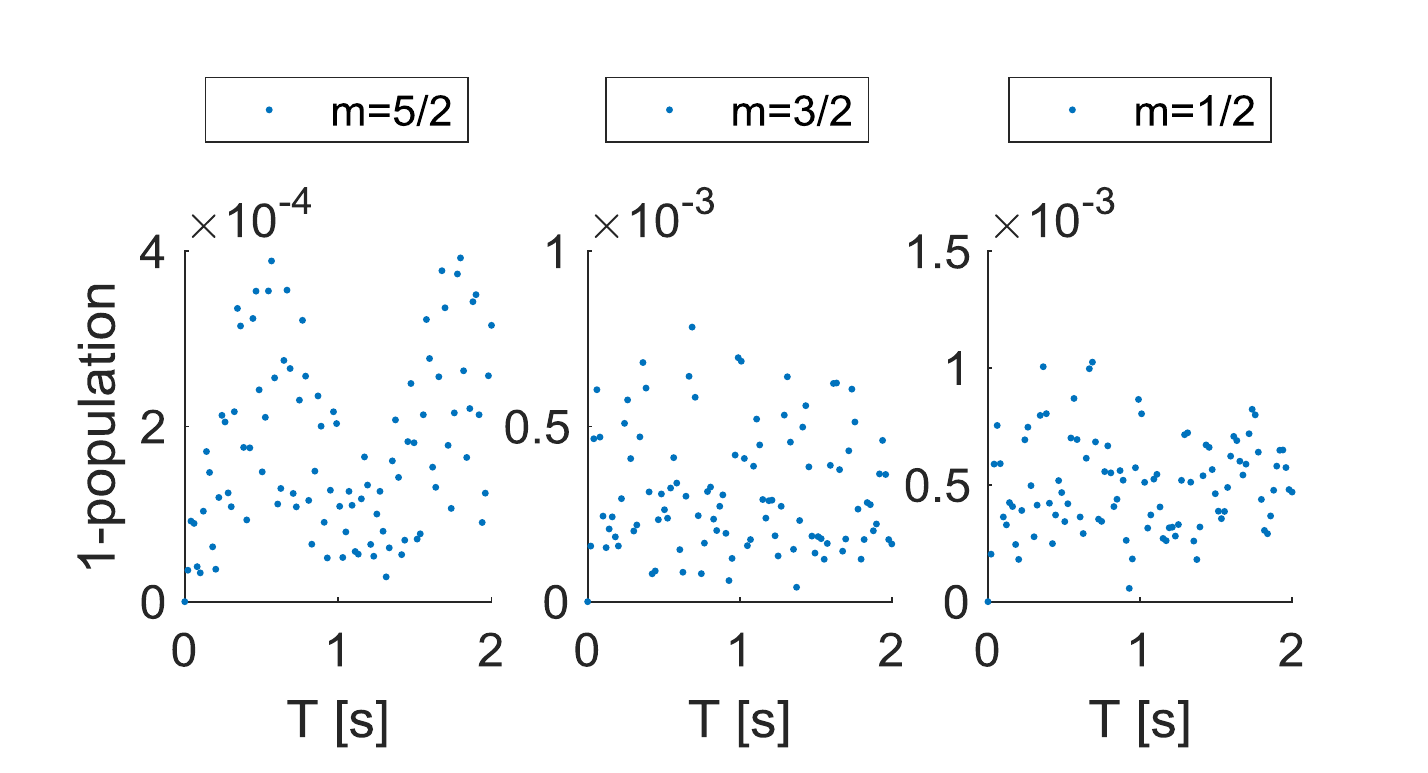}

    (b) \includegraphics[width=0.3\textwidth]{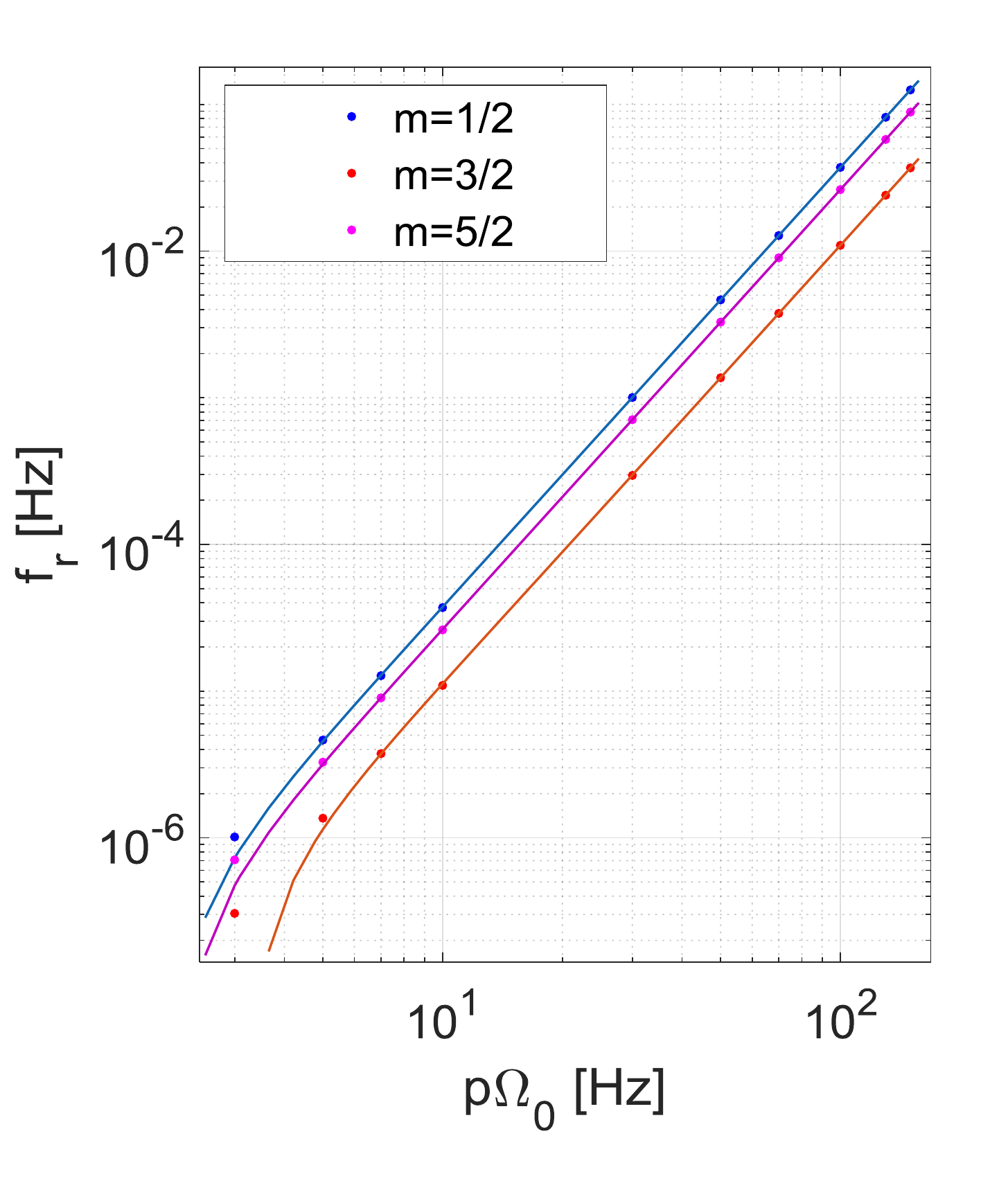}

    \caption{\textbf{Population lost and residual phase in continuous DD sequence.} \textbf{(a)} Population lost for different $m$ initial states. The plots show numerically calculated $1-\left|\left\langle m\right|U_{app}\left(T\right)\left|m \right\rangle\right|^{2}$ as a function of the time $T$, for initial states $\left|m\right\rangle=\left|\frac{5}{2}\right\rangle,\left|\frac{3}{2}\right\rangle,\left|\frac{1}{2}\right\rangle$. \textbf{(b)} Residual frequency $f_{r}$ scaling with $p$, for different initial $m$ states. The plots show slope extracted from of linear fit to $\phi_{act}\left(T\right)-\phi_{app}\left(T\right)$ for $\left|m\right\rangle=\left|\frac{5}{2}\right\rangle,\left|\frac{3}{2}\right\rangle,\left|\frac{1}{2}\right\rangle$ vs different values of $p\Omega_{0}$. The fits are to third order polynomial in $p$. We note that here we show the shift's absolute value. The shift for state $\left|m\right\rangle=\left|\frac{1}{2}\right\rangle$ has opposite sign with respect to the other states.}
    \label{Pop_Lost}
\end{figure}

We calculate the complex amplitudes $\left\langle m\right|U_{act}\left(T\right)\left|m\right\rangle$ and $\left\langle m\right|U_{app}\left(T\right)\left|m\right\rangle$ for interrogation times from $T=0 \text{ s}$ to $T=2 \text{ s}$. In the calculation we set $\Omega_{0}=2\pi\times50 \text{kHz}$.
We would like to verify numerically that indeed $U_{act}\approx1$, since population-leak to other states in the $J$ manifold would result in a loss of contrast in the optical Ramsey fringe. To that aim, we set the conservative values $\delta=Q_{J}=2\pi\times 100\text{ Hz}$, and calculate $1-\left|\left\langle m\right|U_{app}\left(T\right)\left|m \right\rangle\right|^{2}$. The result are shown in Fig. \ref{Pop_Lost}a.
 Fig. \ref{Pop_Lost}a shows that this loss in contrast agrees with the scaling of $p^{2}=\left(\frac{\delta}{\Omega_{0}}\right)^{2}=\left(\frac{Q{J}}{\Omega_{0}}\right)^{2}$, with a pre-factor of order 10. These parameters show that this amplitude deviation is small and can be neglected. We therefore need only to account for the phase acquired by the initial state.
 \begin{figure}[!b]
    \centering
    \includegraphics[width=0.4\textwidth]{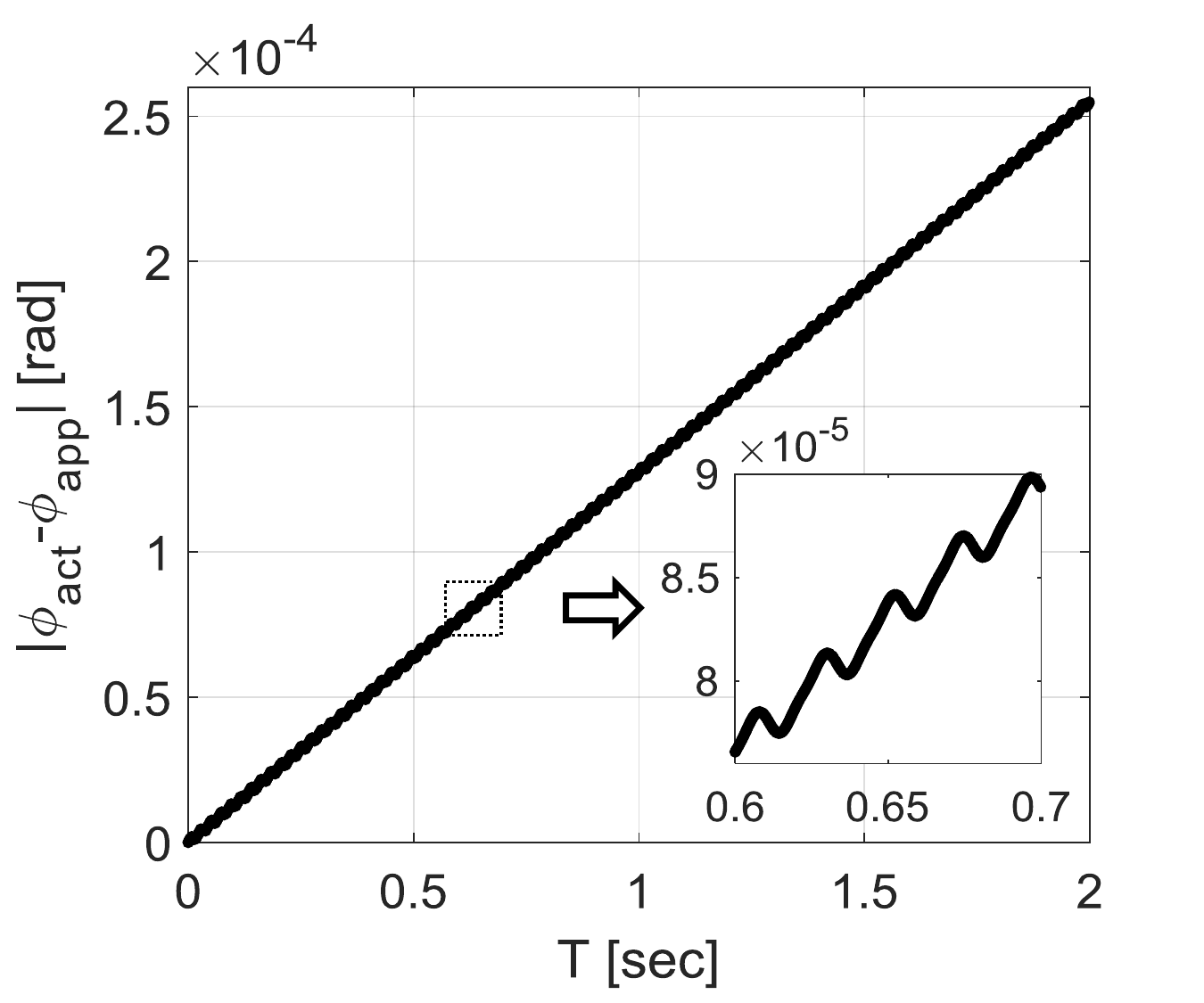}
   \caption{\textbf{Linear and oscillating term in $\phi_{act}-\phi_{app}$.} The residual phase was calculated for the state $\left|\frac{5}{2}\right\rangle$, $Q_{J}=2\pi\times10 \text{ Hz}$, $\delta=2\pi\times1 \text{ Hz}$ and $\Omega_{0}=2\pi\times 50 \text{ kHz}$. A dominant linear trend is shown, with added oscillating term.}
    \label{linear and oscillations}
\end{figure}
We turn to evaluate the phase acquired by the initial state when $U_{act}\left(T\right)$ or $U_{app}\left(T\right)$ are applied. Our figure of merit is $\left|\phi_{act}\left(T\right)-\phi_{app}\left(T\right)\right|$. Again, here we take $\delta=Q_{J}:=p\Omega_{0}$. Using these definitions, residual frequency shift $f_{r}$ arising from the DD sequence corresponds to $\frac{\partial}{\partial T}\left|\phi_{act}\left(T\right)-\phi_{app}\left(T\right)\right|$. This derivative is estimated through a maximum likelihood fit to a linear function. Fig. \ref{Pop_Lost}b shows the scaling of $f_{r}$ as a function of $p$.

These calculation results exhibit a cubic scaling in $p$. As an example, for $p\Omega_{0}=2\pi\times10\text{ Hz}$, the residual shift from the continuous DD part in the sequence is less than $40\mathrm{ \mu Hz}$, which means less than $\approx 9.1\times 10^{-20}$ relative frequency shift.
Next, we verify that the phase-form of a linear term added to an oscillating term is justified. Fig. \ref{linear and oscillations} shows such a behavior for the state $\left|\frac{5}{2}\right\rangle$, $Q_{J}=2\pi\times10 \text{ Hz}$, $\delta=2\pi\times1 \text{ Hz}$ and $\Omega_{0}=2\pi\times 50 \text{ kHz}$. We note, however, that the plot is only qualitative, since it shows a single frequency component, where more frequency components exist. That is due to sample aliasing of the numerical calculation.

We also calculated the expected shift for our DD sequence applied on three ions with two parameters choices, in which $Q_{J}\ne\delta$. The first was using our evaluated experiment parameters: $Q_{J}/2\pi=28,42,28\text{ Hz}$ and $\delta/2\pi=-33,10,53\text{ Hz}$, including our magnetic field gradient and a possible imperfect RF resonance calibration of $10\text{ Hz}$. First, second and third numbers correspond to ion 1 (left), ion 2 (middle) and ion 3 (right). The second was a typical experimental parameters, where $Q_{J}$ is the same as in our experiment and $\delta/2\pi=5,5,5\text{ Hz}$. The resulting residual shift in $\text{mHz}$ is given below.
\begin{center}
\begin{tabular}{ |c|c|c|c| }
\hline
 \multicolumn{4}{|c|}{$m=\frac{5}{2}$} \\
\hline
 & Ion 1 & Ion 2 & Ion 3 \\
\hline
Our experiment parameters &	0.63 &	1.5 &	0.92 \\
\hline
Typical clock parameters &	0.45 &	1.5 & 0.45\\
\hline
\end{tabular}
\end{center}

\begin{center}
\begin{tabular}{ |c|c|c|c| }
\hline
 \multicolumn{4}{|c|}{$m=\frac{3}{2}$} \\
\hline
 & Ion 1 & Ion 2 & Ion 3 \\
\hline
Our experiment parameters &	0.23 &	0.89 &	0.17 \\
\hline
Typical clock parameters &	0.26 &	0.89 & 0.26\\
\hline
\end{tabular}
\end{center}

\begin{center}
\begin{tabular}{ |c|c|c|c| }
\hline
 \multicolumn{4}{|c|}{$m=\frac{1}{2}$} \\
\hline
 & Ion 1 & Ion 2 & Ion 3 \\
\hline
Our experiment parameters &	-0.85 &	-2.4 &	-1.1 \\
\hline
Typical clock parameters &	-0.71 &	-2.4 & -0.71\\
\hline
\end{tabular}
\end{center}

\end{document}